\documentclass[aps,
prd,%
superscriptaddress,%
reprint,%
nofootinbib,%
longbibliography,%
preprintnumbers%
]{revtex4-1}
\pdfoutput=1 
\usepackage[T1]{fontenc} 
\usepackage{hyperref} 
\usepackage{amsmath} 
\usepackage{amsfonts}
\usepackage{amsthm}
\usepackage{amssymb}
\usepackage{graphicx} 
\usepackage{mathrsfs} 
\usepackage{dcolumn} 
\usepackage{color} 
\usepackage{dsfont} 

\newcommand{\mr}[1]{\mathrm{#1}}
\newcommand{\mb}[1]{\mathbf{#1}}
\newcommand{\mc}[1]{\mathcal{#1}}
\newcommand{\ms}[1]{\mathscr{#1}}
\newcommand{\rate}{\tilde{\Gamma}}
\newcommand{\action}{-\log\rate}
\newcommand{\MSbar}{$\overline{\textup{MS}}$}
\newcommand{\muco}[1]{\multicolumn{1}{c}{#1}} 
\newcommand{\Tc}{T_\text{c}}
\newcommand{\yc}{y_\text{c}}
\newcommand{\yp}{y_\text{p}}
\newcommand{\newtext}[1]{#1}
\newcommand{\fixes}[1]{#1}

\graphicspath{{./}{./figs/}}
\newcolumntype{d}[1]{D{.}{.}{#1}}

\newcommand{\Nottingham}{\affiliation{
School of Physics and Astronomy,
University of Nottingham,
Nottingham NG7 2RD,
United Kingdom
}}

\newcommand{\Helsinki}{\affiliation{
Department of Physics and Helsinki Institute of Physics,
PL 64,
FI-00014 University of Helsinki,
Finland
}}

\newcommand{\Marmara}{\affiliation{
Physics Department,
Marmara University,
G\"{o}ztepe Campus,  
34722 Istanbul,
Turkey
}}

\begin{document}


\title{First-order electroweak phase transitions: a nonperturbative update}
\date{May 16, 2022}
\preprint{HIP-2022-10/TH}

\author{Oliver Gould}
\email{oliver.gould@nottingham.ac.uk}
\Nottingham
\Helsinki

\author{Sinan G\"{u}yer}
\email{sinan.guyer@helsinki.fi}
\Marmara
\Helsinki

\author{Kari Rummukainen}
\email{kari.rummukainen@helsinki.fi}
\Helsinki

\begin{abstract}
We study first-order electroweak phase transitions nonperturbatively, assuming any particles beyond the Standard Model are sufficiently heavy to be integrated out at the phase transition. Utilising high temperature dimensional reduction, we perform lattice Monte-Carlo simulations to calculate the main quantities characterising the transition: the critical temperature, the latent heat, the surface tension and the bubble nucleation rate, updating and extending previous lattice studies. We focus on the region where the theory gives first-order phase transitions due to an effective reduction in the Higgs self-coupling and give a detailed comparison with perturbation theory.
\end{abstract}

\maketitle

\section{Introduction}

The electroweak phase transition marks a boundary between epochs in the history of the early universe.
It was the moment in which the Higgs mechanism activated, and the last time in which baryon and lepton number violating processes took place with any degree of rapidity.
In the pure Standard Model (SM) this transition has been accurately studied using a combination of effective field theory and lattice Monte-Carlo simulations \cite{DOnofrio:2015gop}.
The transition has been found to be a crossover, with pseudo-critical temperature $159.5\pm1.5$ GeV, here defined as the temperature at which the susceptibility of the Higgs condensate has a maximum.

Beyond the SM, the nature of the electroweak phase transition depends sensitively on any new physics which lies near the electroweak scale.
Models with a first-order electroweak phase transition can provide the necessary departure from equilibrium for successful baryogenesis at the electroweak scale \cite{Kuzmin:1985mm,Shaposhnikov:1986jp,Shaposhnikov:1987tw,Konstandin:2013caa}.
This possibility also requires new sources of CP violation, and although such sources are strongly constrained by experimental bounds on the electron electric dipole moment \cite{ACME:2018yjb}, viable models are nevertheless possible, e.g.~those based on CP violation in the $\tau$ lepton sector \cite{DeVries:2018aul,Fuchs:2020uoc}.

Collider experiments can in principle test scenarios predicting first-order electroweak phase transitions, because such scenarios require new fields with sizeable couplings to the Higgs field \cite{Ramsey-Musolf:2019lsf}. 
Precise measurements of the Higgs boson’s properties
and interactions constitute a major aim of the Large Hadron Collider (LHC) through its High Luminosity phase \cite{Cepeda:2019klc},
and the construction of a Higgs factory was
deemed the highest priority initiative in the recent European Strategy for Particle Physics update \cite{EuropeanStrategyGroup:2020pow}.

Gravitational wave experiments offer an alternative, and more direct, test of the nature of the electroweak phase transition:
Models in which the electroweak phase transition is first-order also produce a stochastic gravitational wave background, with a spectrum peaked around the mHz range; for reviews see Refs.~\cite{Weir:2017wfa, Hindmarsh:2020hop}.
Planned gravitational wave observatories including LISA \cite{Audley:2017drz, Caprini:2015zlo, Caprini:2019egz}, DECIGO \cite{Kawamura:2011zz}, BBO \cite{Harry:2006fi} and Taiji \cite{Guo:2018npi} may be sensitive to this signal.

For a given particle physics model, predicting the thermal evolution of the universe around the electroweak phase transition requires tackling a number of theoretical challenges.
At high temperature, infrared bosonic degrees of freedom become highly occupied, thereby increasing their effective couplings.
This results in large theoretical uncertainties at low perturbative orders, often amounting to several orders of magnitude undertainty for the gravitational wave peak amplitude \cite{Croon:2020cgk, Gould:2021oba}.
Sufficiently long wavelength modes become strongly coupled.
For non-Abelian gauge theories, such as the electroweak theory, this happens generically, and is known as Linde's Infrared Problem \cite{Linde:1980ts}.
It implies that any perturbative approach to the study of the electroweak phase transition is fundamentally incomplete.

Direct lattice simulation of the electroweak theory at high temperatures is thwarted by the Nielsen-Ninomiya theorem \cite{Nielsen:1980rz, Nielsen:1981xu}, which prevents the simulation of Weyl fermions with a chiral gauge coupling.
Nevertheless, a nonperturbative solution to Linde's Infrared Problem is possible, and was formulated in Refs.~\cite{Kajantie:1993ag, Farakos:1994xh, Kajantie:1995kf}.
It involves first constructing a three-dimensional effective field theory (3d EFT) for the nonperturbative infrared bosonic modes, and then simulating this EFT on the lattice.
In the first step, called {\em high temperature dimensional reduction}, the effects of all weakly coupled modes are accounted for perturbatively, including those of the chiral fermions.
The result is a much simpler EFT describing only the infrared bosonic modes, which can be straightforwardly simulated on the lattice.
This is the approach we will adopt in the following.

To study time-dependent and non-equilibrium quantities, such as the bubble nucleation rate, one must go beyond dimensional reduction which describes only static quantities.
Direct lattice simulations are then thwarted by the sign problem of real-time evolution.
However, at high temperatures, an alternative EFT approach is possible: The time evolution of the longest wavelength modes, together with screening and other plasma effects, is described by Langevin-type equations up to corrections suppressed by $1/\log(1/g)$ \cite{Bodeker:1998hm}.
This framework was adopted in Ref.~\cite{Moore:2000jw} to nonperturbatively compute the bubble nucleation rate, and is the approach we will adopt in this paper.

A wide range of theories beyond the Standard Model (BSM) have the same infrared degrees of freedom in the vicinity of the electroweak phase transition, and are therefore described by the same high temperature EFT as the SM.
Any new BSM particles need only be heavy compared to the effective mass of the Higgs, which becomes light in the vicinity of the transition.
Only the infrared degrees of freedom may have nonperturbative effective couplings, so only these require numerical lattice simulations.
In this paper we will perform new lattice simulations to provide a quantitatively reliable description of the thermodynamics of the SM-like high temperature EFT, extending the results of Refs.~\cite{Kajantie:1995kf, Rummukainen:1998as, Moore:2000jw}.
We will focus on regions of parameter space where the electroweak phase transition is of first order.
This can occur due to Higgs-portal couplings, through which BSM particles can induce the effective Higgs self-coupling to be weaker than it is in the SM.

This approach has been successfully utilised to study phase transitions in several BSM theories, including the Minimally Supersymmetric Standard Model (MSSM) \cite{Laine:1996ms, Cline:1996cr, Farrar:1996cp, Cline:1997bm}, the two Higgs doublet model (2HDM) \cite{Losada:1996ju, Andersen:1998br, Andersen:2017ika, Gorda:2018hvi}, the singlet extended Standard Model (xSM) \cite{Brauner:2016fla, Gould:2019qek}, and the real triplet scalar extension of the Standard Model ($\Sigma$SM) \cite{Niemi:2018asa}.
In Ref.~\cite{Gould:2019qek} lattice simulations of the SM-like high temperature EFT were used to provide the first nonperturbative predictions of the gravitational wave signal of a first-order cosmological phase transition.
These were however limited due to the existence of only a single nonperturbative calculation of the bubble nucleation rate \cite{Moore:2000jw}.
One motivation for this paper was to remedy this deficit by nonperturbatively computing the bubble nucleation rate at other parameter points, and thereby extending the scope for making nonperturbative predictions of the gravitational wave signal of first-order electroweak phase transitions.

A final key motivation for this paper is the delineation of the validity of perturbation theory to describe first-order electroweak phase transitions; see also Ref.~\cite{Kainulainen:2019kyp}.
While perturbation theory is fundamentally incomplete at high temperatures due to the Infrared Problem, it is nevertheless still useful when the transition is strongly first order.
In this case, one can construct the first few orders of a perturbative expansion in ratios of couplings.
This expansion contains half-integer \cite{Kapusta:1979fh, Arnold:1992rz} and quarter-integer \cite{Ekstedt:2020abj, Ekstedt:2022zro, Ekstedt:2022ceo} powers of the gauge coupling.
It therefore both converges slowly and breaks down at finite order.
Its quantitative reliability can only be reliably determined by a nonperturbative calculation.

In Sec.~\ref{sec:background} we discuss the high temperature EFT of the electroweak phase transition, giving an overview of our existing knowledge of this EFT and of its applications to BSM theories.
In the following sections we present our calculations and results for the thermodynamics of the EFT, starting with the equilibrium thermodynamics in Sec.~\ref{sec:equilibrium}.
In Sec.~\ref{sec:nucleation} we present our calculations and results for the bubble nucleation rate.
Though requiring significant computational resources, we are able to take controlled continuum limits for the first time, and to study the parametric dependence of the bubble nucleation rate.
In Sec.~\ref{sec:bubbles} we show some visualisations of critical bubbles taken from our lattice simulations, extended into a movie of bubble growth which can be seen at \cite{videos}.
In Sec.~\ref{sec:cosmo} we utilise our nucleation rate results within a cosmological context.
We conclude in Sec.~\ref{sec:conclusions}.
A number of appendices collect details of our continuum extrapolations and perturbative results.

\section{Electroweak physics at high temperatures} \label{sec:background}

At high temperatures, the low energy modes of bosonic fields become highly occupied.
The thermodynamics of these modes is captured by a classical 3d EFT.
For the electroweak phase transition, the relevant EFT contains all the bosonic fields of the SM: the Higgs, and the gauge bosons.
In the presence of additional BSM bosonic fields, these fields may or may not enter the EFT, depending on their thermal effective masses.

At a technical level, the 3d nature of the high-temperature EFT can be understood from the imaginary time formalism, whereby thermodynamics is formulated as quantum field theory on $\mathbb{R}^3\times\mathrm{S}^1$.
The circle $S^1$ has circumference $1/T$ and is referred to as the imaginary time direction.
Quantum fields are then expanded in Fourier modes of the imaginary time direction with frequencies $n \pi T$, where $n$ is an even integer for bosons and an odd integer for fermions.
These are called Matsubara (or Kaluza-Klein) modes.
At sufficiently high temperatures, such that $\pi T$ is large compared to other energy scales, all nonzero ($n\neq 0$) Matsubara modes become heavy and decouple from the physics of the zero ($n=0$) modes.
The zero modes are constant in the imaginary time direction; the EFT that describes them is therefore 3d.

This EFT describes the static long-wavelength modes of the thermal bath, in particular those with energies up to $O(gT)$.
Of these modes, the Debye-screened temporal gauge fields are heaviest, and in turn can be integrated out \cite{Farakos:1994kj, Braaten:1995jr, Kajantie:1995dw}. Any BSM modes in this energy range can also be integrated out at this stage; see for example Ref.~\cite{Gorda:2018hvi, Gorda:2018hvi}. The result is a simpler 3d EFT, containing only the spatial, or magnetic, components of the gauge fields, as well as any sufficiently light scalar fields.

In principle this EFT contains the full complement of $\mr{SU}(3)\times \mr{SU}(2)\times \mr{U}(1)$ spatial gauge fields of the Standard Model. However, couplings between the Higgs and the $\mr{SU}(3)$ gauge fields of the EFT are both indirect and parametrically suppressed, and hence the latter can be dropped if we are only interested in the electroweak phase transition. In addition, we choose to drop the $\mr{U}(1)$ hypercharge gauge field. This is for two reasons: first, due to the smallness of the Weinberg angle the $\mr{U}(1)$ gauge field has only a numerically small effect on the phase transition; second, due to the absence of $\mr{U}(1)$ gauge self-interactions, this field does not suffer from Linde's Infrared Problem, and consequently perturbation theory provides a reliable guide to its contributions \cite{Kajantie:1996qd}.

In this paper, we study the 3d SU(2) gauge-Higgs theory, defined by the following Lagrangian density,
\begin{align}
 \ms{L}_3 &= \frac{1}{4}F_{ij}^a F_{ij}^a
 + D_i\phi^\dagger D_i\phi 
 \nonumber \\
 &\quad
 + (m_3^2 + \delta m_3^2) \phi^\dagger \phi
 + \lambda_3 (\phi^\dagger \phi)^2
 \;, \label{eq:lagrangian}
\end{align}
where we have defined
\begin{align}
 F_{ij}^a &= \partial_i A_j^a - \partial_j A_i^a + g_3 \epsilon^{a b c} A_i^b A_j^c,
 \;, \label{eq:Fij} \\
 D_i \phi &= \partial_i \phi - \frac{i}{2} g_3 \sigma^a A_i^a \phi
 \;, \label{eq:Dphi}
\end{align}
the letters $i, j$ run over the spatial indices, $a, b, c$ run over the indicies of the adjoint representation of SU(2), $\epsilon^{abc}$ is the Levi-Civita symbol and $\sigma^a$ are the Pauli matrices.
Here the Higgs field $\phi$ lies in the fundamental representation of SU(2)
with gauge coupling $g_3$, and we have left the corresponding indices implicit.

As all couplings have nonzero mass dimension, there can only be nontrivial dependence on the dimensionless ratios,
\begin{align}
x(T) &\equiv \frac{\lambda_{3}}{g_3^2},
&
y(T) &\equiv \frac{m_{3}^2}{g_3^4}. \label{eq:xy}
\end{align}
Including the $U(1)$ hypercharge gauge field, with gauge coupling $g_3^{'2}$, would introduce an additional dimensionless parameter $z(T)=g_3^{'2}/g_3^2$.

This 3d EFT \eqref{eq:lagrangian} has been studied extensively both perturbatively and nonperturbatively in Refs.~\cite{Farakos:1994kx, Farakos:1994xh, Buchmuller:1994qy, Kajantie:1995dw, Kajantie:1995kf, Rummukainen:1998as, Moore:2000jw, Moore:2014dsa, York:2014ada}.
The framework for lattice simulations was initially developed in Ref.~\cite{Farakos:1994xh}, and exact and $O(a)$ improved lattice-continuum relations were derived in Refs.~\cite{Laine:1995np,Laine:1997dy,Moore:1997np}.
Efficient multicanonical Monte-Carlo algorithms for the study of first-order phase transitions in this theory were presented in Ref.~\cite{Kajantie:1995kf} (see also Ref.~\cite{Laine:1998qk}).

The phase diagram of the theory contains a line of first-order phase transitions $y=\yc(x)$, for which $x\ll 1$.
This line ends in a critical point $(x,y)=(x_*,y_*)$ at which there is a second-order phase transition in the 3d Ising universality class \cite{Gurtler:1997hr, Rummukainen:1998as, Laine:1998jb}.
Numerically, this happens at
\begin{align}
x_* &= 0.0983(15),
&
y_* &= -0.0175(13),
\end{align}
for \MSbar\ renormalisation scale $\mu_3=g_3^2$ (which we use throughout).
At other renormalisation scales, the position of this point is modified as
\begin{align}
y_* \to y_* + \frac{1}{(4\pi)^2}\left(-\frac{51}{16}-9x^2+12x^2\right)\log\frac{\mu_3}{g_3^2}.
\end{align}
This identity is exact, due to the superrenormalisability of the theory~\cite{Farakos:1994kx}, and $x_*$ is unmodified.

To better understand how the parameters of the 3d EFT relate to those of the full ultraviolet theory, let us briefly consider an example: the SM plus an additional real scalar field (xSM).
Denoting the additional field by $\sigma$, the following Higgs portal couplings will generically be present
\begin{align}
\Delta \ms{L} = -\frac{1}{2}a_1\sigma \phi^\dagger \phi - \frac{1}{2}a_2\sigma^2\phi^\dagger\phi.
\end{align}
These terms modify the 3d effective couplings away from their SM values.
At tree level in the xSM, the 3d coupling $x$ reads \cite{Brauner:2016fla, Niemi:2021qvp}
\begin{align} \label{eq:xxSM}
x = \frac{m_H^2}{8m_W^2}\left(1 + \frac{m_\sigma^2-m_H^2}{m_H^2}\sin^2\theta \right),
\end{align}
where $m_H$ and $m_W$ are the physical (pole) masses of the Higgs and $W$-boson, $m_\sigma$ is the physical mass of the BSM scalar particle and $\theta$ is its mixing angle with the Higgs.
The mixing angle is generically nonzero for $a_1\neq 0$.

In the pure SM, $x\approx m_H^2/8m_W^2 \approx 0.3 > x_*$ and the electroweak phase transition is a crossover \cite{DOnofrio:2015gop}.
In the xSM, the second term in Eq.~\eqref{eq:xxSM} can reduce the value of $x$, if $m_\sigma<m_H$, and therefore move the electroweak phase transition towards being first order.
In fact, this possibility is favoured by the recent measurement of the $W$ boson mass by the CDF collaboration \cite{CDF:2022hxs, Lopez-Val:2014jva}.

In the presence of relatively large portal couplings, one-loop corrections can also modify the value of $x$ significantly, especially when the real scalar is not too heavy.
In the $Z_2$-symmetric limit of the xSM the scalar fields do not mix, $\sin\theta=0$, and the leading corrections from the real scalar field arise at one loop.
Assuming the thermal effective mass of the real scalar lies in the range $gT \lesssim m_{\sigma, 3} \lesssim \pi T$, this correction to $x$ is of order
\begin{equation}
\Delta x  \sim \frac{a_2^2 T}{\pi g^2 m_{\sigma, 3}}
\end{equation}
where $g$ is the weak gauge coupling.
This correction is also generically negative for thermal effective masses of order $gT$.
As a numerical example, for a $Z_2$-symmetric real scalar with mass $m_\sigma= 200$~GeV, one finds $x<x_*$ for $a_2\gtrsim 1.3$ \cite{Gould:2019qek}.

Current experimental bounds still leave ample room for the value of $x$ to be modified.
This is because $x$ is determined by the Higgs four-point self-coupling, and, while the Higgs mass has been relatively precisely measured, $m_H=125.25\pm 0.17$~GeV~\cite{Zyla:2020zbs}, its self-couplings are much less well constrained.
Measurements of Higgs pair production have be used to directly constrain the Higgs three-point coupling.
The current limit can be expressed as $\lambda/\lambda_{\rm SM} \in (-5.0,12.0)$ at 95\% confidence~\cite{ATLAS:2019qdc}.
Direct constraints on the four-point Higgs coupling are even weaker.
Such weak constraints leave open the possibility that $x<x_*$, and hence for a first-order electroweak phase transition.
This will be the focus of the current article.

\section{Equilibrium quantities} \label{sec:equilibrium}

The equilibrium thermodynamics of the 3d EFT is relatively straightforward to study on the lattice.
There is no true order parameter in the $SU(2)$ Higgs theory, in the sense of the Landau theory of continuous phase transitions \cite{Landau:1937obd}:
The naive candidate $\langle \phi \rangle$ is equal to zero for all temperatures, being gauge dependent \cite{Elitzur:1975im}.
The volume average of the quadratic scalar condensate, $\langle \phi^\dagger \phi \rangle$, provides a means to distinguish the different phases, although it is positive for all temperatures.
It plays a role analogous to the density in liquid-gas transitions.

\subsection{Critical temperature} \label{sec:critical_temperature}

As outlined in Sec.~\ref{sec:background}, the phase diagram of the 3d $SU(2)$ Higgs theory consists of a line of first-order phase transitions $\yc(x)$ for $x\in (0,x_*)$, terminating in a second-order phase transition at $x=x_*$.
For larger values of $x$ the transition is merely a crossover.

A given 4d BSM theory at a given temperature maps to a point on the space of couplings of the 3d EFT.
As the temperature varies, the thermodynamics of the 4d theory trace out a trajectory through the couplings of the 3d EFT, parametrised by the temperature $(x(T),y(T),g_3^2(T))$.
There is a first-order phase transition if this trajectory intersects the surface of first-order phase transitions, $y=\yc(x)$.

\begin{figure}[t]
    \centering
    \includegraphics[width=0.48\textwidth]{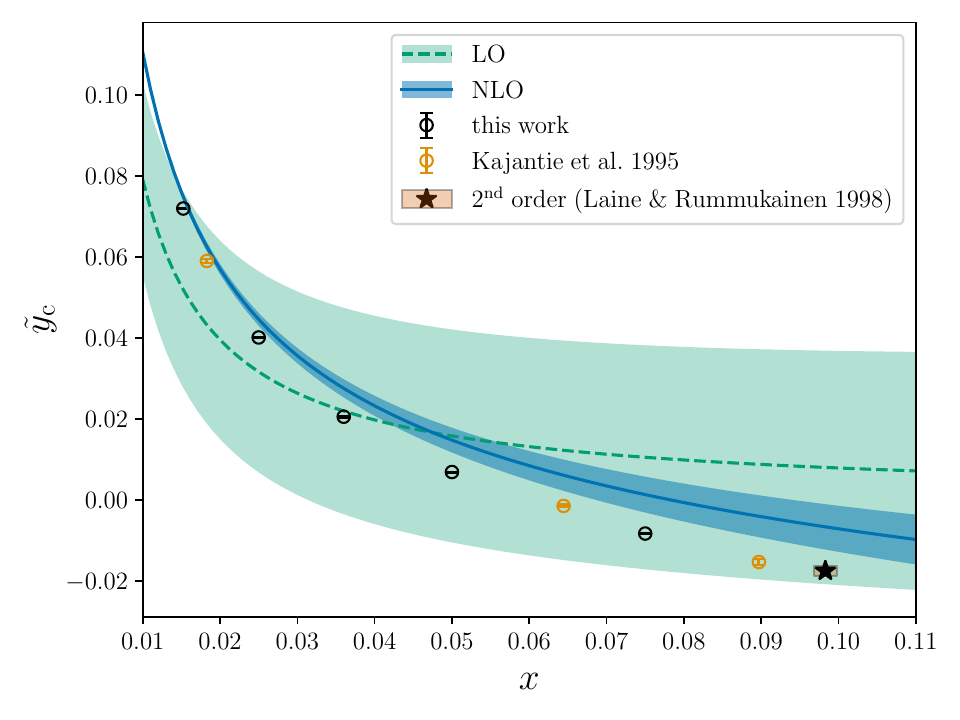}
    \caption{The phase diagram, showing the line of first-order phase transitions and \newtext{the second order phase transition at} its endpoint\newtext{, marked with a star}.
    Lattice results from this paper and from Refs.~\cite{Kajantie:1995kf, Laine:1998jb} are plotted together with two perturbative approximations.
    \newtext{Note that the perturbative approach fails to predict an endpoint to the line of first-order phase transitions.}
    The renormalisation group invariant quantity $\tilde{y}_\text{c}\equiv \yc - \beta_y \log(\mu_3/g_3^2)$ has been plotted (see Eq.~\eqref{eq:running_y}), rather than simply $\yc$.
    The green and blue bands for the perturbative results reflect the renormalisation scale dependence as it is varied over $\mu_3/g_3^2\in [\tfrac{1}{\sqrt{10}},\sqrt{10}]$.}
    \label{fig:yc}
\end{figure}

The speed at which the trajectory moves through the space of 3d couplings, or equivalently the tangent vector, is naturally quantified by
\begin{align} \label{eq:etas}
\eta_x &\equiv \frac{dx}{d\log T}, &
\eta_y &\equiv \frac{dy}{d\log T}, &
\eta_{g_3^2} &\equiv \frac{dg_3^2}{d\log T}.
\end{align}
The generic structure of dimensional reduction implies that $\eta_x \sim g^2 \ll \eta_y \sim g^{-2}$, where $g^2$ denotes a perturbative coupling \cite{Kajantie:1995kf, Gould:2019qek}.
Thus, for any perturbative 4d model, the trajectory is almost parallel to the $y$ axis.
Note also that $\eta_{g_3^2}= g_3^2\cdot(1+O(g^2))$, and hence the relation between units in 3d and 4d tracks the temperature up to small corrections.

\begin{figure}[t]
    \centering
    \includegraphics[width=0.48\textwidth]{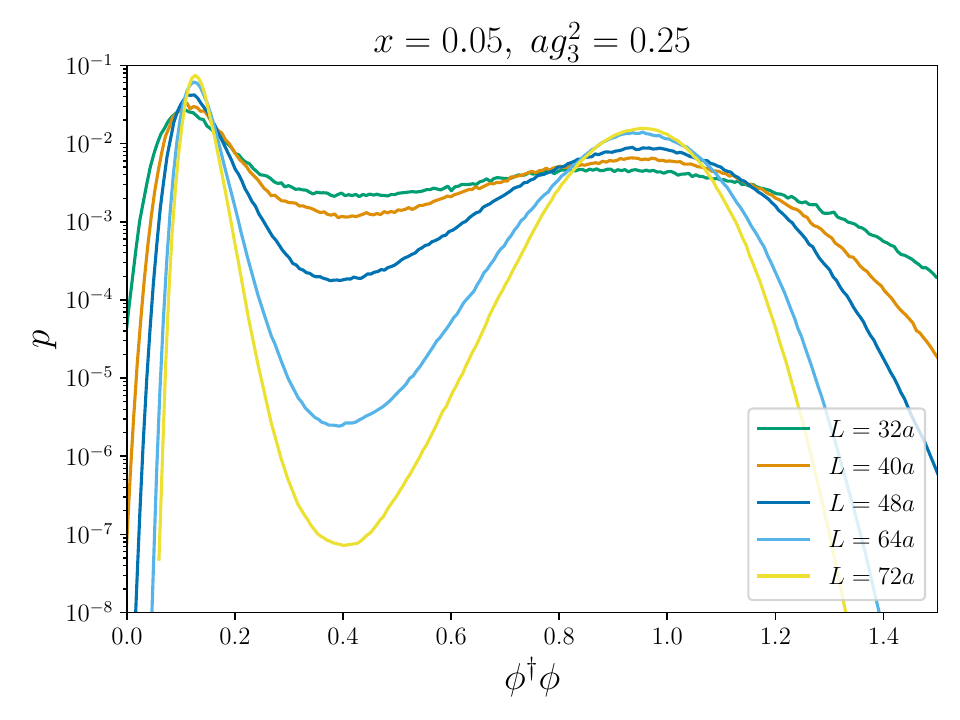}
    \includegraphics[width=0.48\textwidth]{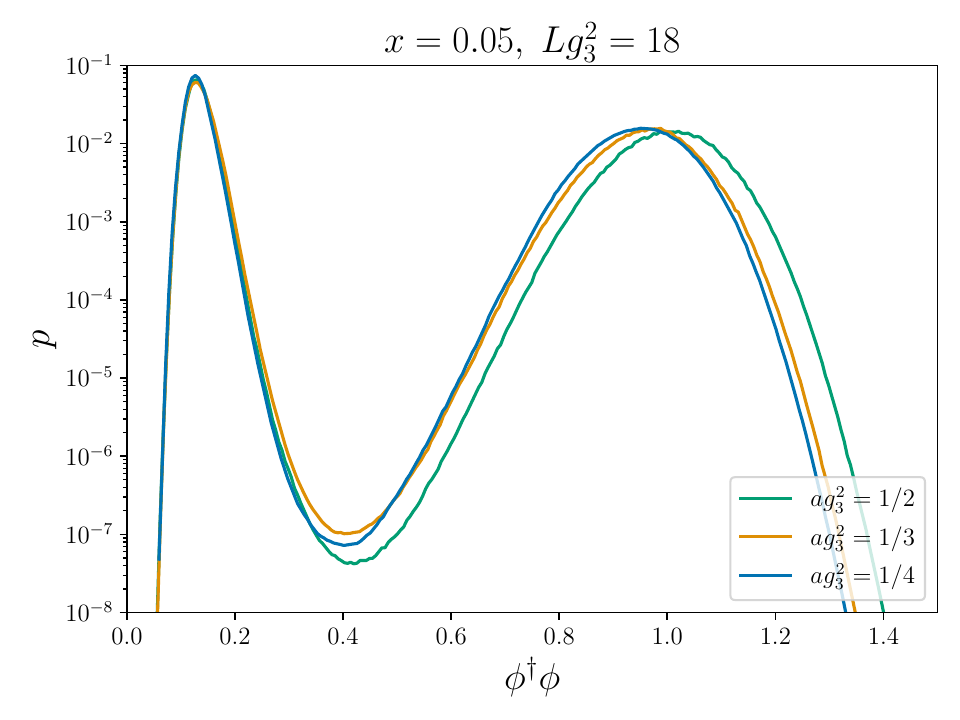}
    \caption{Histograms of the spatially averaged Higgs quadratic condensate, renormalised to match the \MSbar\ scheme in the continuum limit.
    The two peak structure demonstrates the coexistence of phases, and the first-order nature of the transition.
    The upper panel shows the physical volume dependence of the histogram, from which one can see that states between the two phases become exponentially unlikely as the volume $L^3$ grows.
    \newtext{The Gaussian width of each peak scales as $1/\sqrt{V}$ \cite{Farakos:1994xh}.}
    The lower panel shows the relatively mild lattice spacing ($a$) dependence of this quantity.}
    \label{fig:hists}
\end{figure}

To calculate $\yc(x)$, we fix a value of $x<x_*$ and vary $y$.
At large positive (negative) values of $y$, equivalent to high (low) temperatures, there is a single phase, the symmetric (broken) phase.
In between these two extremes, there is a region of $y$ for which both phases coexist.
At the critical temperature, $y=\yc$, the two phases have equal free energies, and hence equal probabilities in the thermal ensemble: a histogram of $\phi^\dagger \phi$ will show two peaks of equal probability; see Fig.~\ref{fig:hists}.

An efficient approach to calculate the critical temperature makes use of an exact relation between the probability distribution of $\phi^\dagger \phi$ at different values of $y$ \cite{Ferrenberg:1988yz, Ferrenberg:1989ui}.
We denote this probability distribution, evaluated at some value $\varphi^2$, by
\begin{align} \label{eq:probability_path_integral}
p_{y}(\varphi^2) &\propto
\int \mc{D}\phi \mc{D}A\ \delta \left(\varphi^2 V - \int_\mb{x} \phi^\dagger \phi \right) e^{-\int_\mb{x} \ms{L}_3},
\end{align}
where $\delta$ is the Dirac delta function, $\int_\mb{x}$ denotes integration over space, and $V$ is the spatial volume.
The crucial relation, which follows from the definition, is
\begin{align} \label{eq:reweighting}
p_{y'}(\varphi^2)
=
e^{-(y'-y)\varphi^2 V}
p_{y}(\varphi^2).
\end{align}
Use of this relation is referred to as \textit{reweighting}.
In principle one can reweight the results of a simulation at a single value of $y$ to determine the probability distribution at all values of $y$.
In practice, there is an overlap problem if one tries to reweight too far, which will manifest through larger statistical errors.
Once one has found an approximate value for $\yc$, one can reweight to find a very much more accurate value.
A corresponding relation exists for the probability distribution of $(\phi^\dagger\phi)^2$ and changes in $x$, though we have not made use of it.

The two-peak histograms of Fig.~\ref{fig:hists} have been reweighted to the critical point, at which the probability in each phase (area under each peak) is equal.
The boundary between phases can be taken as the local minimum of the probability, though the precise choice of boundary is unimportant: in large lattices the value of $\yc$ attained is exponentially insensitive to this choice.
Fig.~\ref{fig:hists} shows examples of the volume and lattice-spacing dependence on a range of cubic lattices.
\newtext{For each physical parameter point, simulations were carried out for at least 4 lattice spacings, and for each of these, between 3 and 6 different volumes were simulated; see Table~\ref{table:equilibrium_lattices}. Suitable choices for lattice spacings and volumes can be gleaned from previous work \cite{Kajantie:1995kf}, and from comparing results for different lattices.}

As the theory is (nonperturbatively) gapped, for sufficiently large volumes the volume dependence \newtext{of bulk observables, such as $\int_\mathbf{x}\phi^\dagger\phi$ in a given phase,} is exponentially suppressed.
\newtext{This is because the exponential barrier isolates the two phases, ensuring that the thermodynamics of one phase is independent of the presence of the other.
The volume dependence of $\yc$ is more subtle, as it depends on both phases, and in fact on the precise definition of $\yc$ at finite volume. Defining $\yc$ through the maximum of the susceptibility, or the minimum of the Binder cumulant, leads to $1/V$ dependence \cite{borgs1990rigorous, Kajantie:1995kf}, because the relative widths of the two peaks play a role.
So too does defining $\yc$ as the temperature at which the two peaks of the histogram (Fig.~\ref{fig:hists}) have equal height \cite{Kajantie:1995kf}.
However, defining $\yc$ as the temperature at which the two peaks of the histogram have equal probability leads to exponentially suppressed volume corrections \cite{borgs1990rigorous, borgs1992finite, janke1993accurate}. This is the definition we adopt.}
Extrapolations to infinite volume were therefore carried out simply by averaging the results of the largest lattices, ensuring that they agree within error.
Following this, extrapolations to zero lattice spacing were performed with polynomial fits of the lowest degree such that $\chi^2/\text{d.o.f.}\sim 1$, starting with $1+a$.
Some examples of our continuum extrapolations for $\yc$ are shown in Appendix~\ref{appendix:extrapolations}.
The full dataset can be found at \cite{data}.

Errors for individual lattices were estimated by performing a jackknife resampling, after blocking the data into ten blocks, as we do throughout.
For continuum extrapolations, we quote the error on the fit parameter.

We have also made use of an alternative method to calculate the critical temperature.
In lattices with one side much longer than the other, which we refer to as \textit{cylindrical} lattices, the local minimum in the probability distribution becomes wider and flatter as the ratio between long and short sides grows.
Physically this wide, flat region in the probability distribution is dominated by configurations where the two phases coexist.
The tension of the phase boundary ensures that the boundary will line up perpendicularly to the long axis, as other configurations have much higher free energy.
At the critical temperature, the free energy (and hence probability) of such a configuration is independent of the precise fraction of broken phase versus symmetric phase in the lattice.
This is because the free energy densities of both phases are the same, and because both configurations have the same area of phase boundaries.
Slightly away from the critical temperature, changing the fraction of broken phase will change the free energy, thus the probability distribution of the order parameter will no longer be flat.
One can thus tune $y$ to the critical temperature by reweighting the probability distribution  until the region between phases becomes flat.
A related method was used in Ref.~\cite{Moore:2019lua}.
An advantage of this method is that there is no need to wait for slow tunnellings between phases, as one only needs to know the probability distribution for mixed configurations.
However, a disadvantage is that one must first find the appropriate range over which the probability distribution becomes flat, and this is not known \textit{a priori}.
The two methods we have used for calculating $\yc$ agree within error.
\newtext{Some example plots showing} this can be found in Appendix~\ref{appendix:extrapolations}.

Our continuum-extrapolated results for $\yc$ are presented in Fig.~\ref{fig:yc}, together with previous results from the literature.
Altogether the data is now rather dense in $x$, exposing the smooth behaviour of the function $\yc(x)$.
The numerical values are collected in Table \ref{table:results_yc}.

While Linde's Infrared Problem renders perturbation theory inherently incomplete, it is nevertheless possible to compute the first few orders of an expansion in $x$.
With a two-loop computation one can determine the leading order (LO) and next-to-leading order (NLO) behaviour in $x$.
This is outlined in Appendix \ref{appendix:perturbative_results}.
The calculation of yet higher order terms involves infinite sets of diagrams within the 3d EFT, which we do not attempt here; see Refs.~\cite{Ekstedt:2022zro, Ekstedt:2022ceo}.
However, by scaling arguments one can determine that pure gauge diagrams yield an expansion in powers of $x$, while pure Higgs diagrams yield an expansion in powers of $x^{3/2}$.
The expansion for the full theory is thus a dual expansion in powers of $x$ and $x^{3/2}$, up to logarithms, thereby motivating the following fit function \cite{Kajantie:1997tt}
\begin{align} \label{eq:yc_fit}
\yc^{\rm fit}(x) &= \frac{1}{128\pi^2 x}\bigg[
1
+x\left(\frac{63}{2} \log\frac{3}{2}-\frac{33}{4}-\frac{51}{2} \log 8 \pi  x\right)
\nonumber \\
&\quad
+c_{3/2}x^{3/2}
+c_2 x^2
+c_{5/3} x^{5/2}
\bigg].
\end{align}
Performing a least-squares fit, we find
\begin{align}
c_{3/2}&=16(6),& c_2&=-490(60),&  c_{5/2}&=980(150),
\end{align}
with $\chi^2/\text{d.o.f.}=6$.
The relatively large values of the expansion coefficients suggest that perturbation theory in $x$ breaks down around $x \sim 0.05$.
Note that despite the breakdown of perturbation theory, the coefficients $c_{3/2}$, $c_2$ and $c_{5/2}$ (but not the coefficients of higher powers of $x$) can in principle be computed much more accurately within perturbation theory, with a resummed 3-loop computation.
If this is performed, the fit to the lattice data should be updated to include higher powers of $x$, or an alternative functional form which describes the data better at larger $x$.

\subsection{Latent heat} \label{sec:latent_heat}

The latent heat $L$ is the change in enthalpy density between phases at the critical temperature.
It gives a measure of the energy per unit volume released during the phase transition.
In the cosmological evolution, some degree of supercooling to temperatures below the critical temperature will take place, due to the slowness of bubble nucleation in comparison to Hubble expansion (see Sec.~\ref{sec:cosmo}).
The resulting energy released is expected to be greater than $L$, as additionally energy will be released proportional to the free energy density difference $\Delta f$ between phases.
The latent heat nevertheless gives a bound on the energy released, and a reliable estimate for small supercooling.

The latent heat is determined by the rate of change of the free energy difference between phases, evaluated at the critical temperature
\begin{align}
\frac{L}{\Tc^4} \equiv \frac{d}{d\log T}\left(\frac{\Delta f}{T^4}\right)\bigg|_{\Tc}.
\end{align}
Using the chain rule, and that $\Delta f = 0$ at the critical temperature, this can be written as
\begin{align} \label{eq:latent_heat_condensates}
\frac{L}{\Tc^4} = \frac{g_3^6}{\Tc^3}
\left(
\eta_y \Delta \langle \phi^\dagger \phi\rangle_{\rm c}
+ \eta_x \Delta \langle (\phi^\dagger \phi)^2\rangle_{\rm c}
\right),
\end{align}
where the condensates are expressed in units of $g_3^2$\newtext{, and the eta functions are defined in Eq.~\eqref{eq:etas}.}
The condensates are defined in terms of the derivatives of the free energy density with respect to the renormalised couplings \cite{Farakos:1994xh},
\begin{align}
\Delta \langle \phi^\dagger \phi\rangle &\equiv \frac{1}{g_3^6} \frac{\partial}{\partial y}\left(\frac{\Delta f}{T} \right), \\
\Delta \langle (\phi^\dagger \phi)^2\rangle &\equiv \frac{1}{g_3^6} \frac{\partial}{\partial x}\left(\frac{\Delta f}{T} \right).
\end{align}
We have scaled out powers of $g_3^2$ so that the condensates are dimensionless.
The quantities then correspond more closely to what is measured on the lattice, in which we adopt units where $g_3^2=1$.

These definitions for the condensates, being the derivative of a finite quantity with respect to finite couplings, yield the necessary counterterms to make the condensates finite.
The renormalised quartic condensate is equal to a linear combination of the unrenormalised quartic and quadratic condensates, the latter arising as a consequence of the mass counterterm depending on $x$.
Note that the condensate corresponding to the partial derivative with respect to $g_3^2$ (at fixed $x$ and $y$) is equal to zero at the critical temperature, being proportional to $\Delta f$.

The natural hierarchy $\eta_y \gg \eta_x$ suggests that the second term in Eq.~\eqref{eq:latent_heat_condensates} can be neglected.
For relatively weak transitions, this is indeed correct.
However, in very strong transitions the quartic condensate can become larger than the quadratic condensate, so in the following we will study both terms.

\begin{figure}[t]
    \centering
    \includegraphics[width=0.48\textwidth]{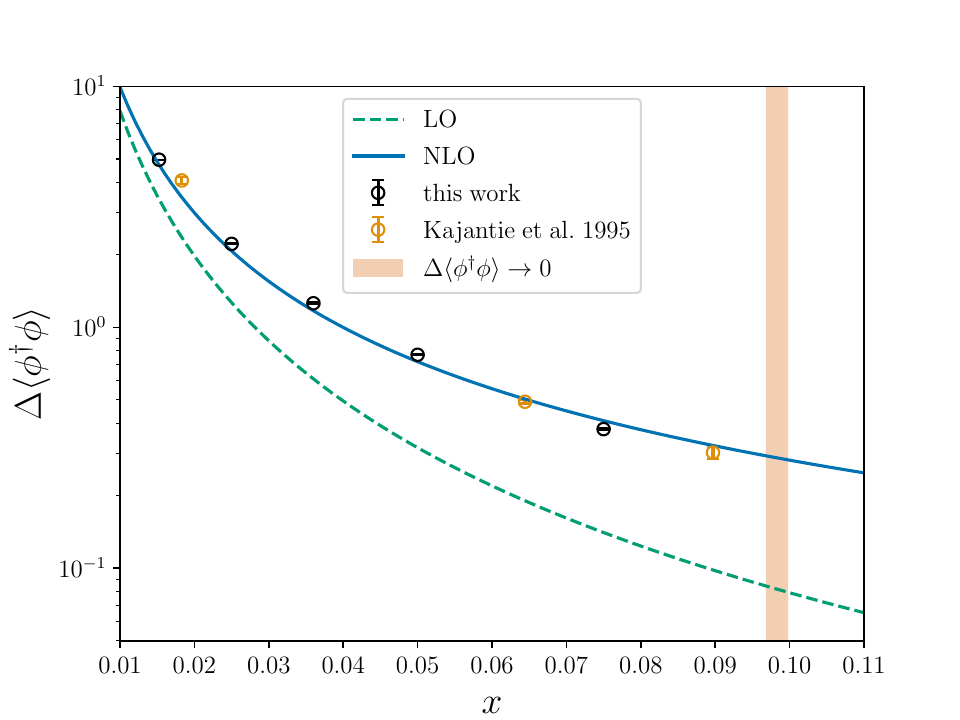}
    \caption{
    The discontinuity in the quadratic Higgs condensate.
    Note that this goes to zero at $x_*=0.0983(15)$, illustrated as the vertical orange band.
    Lattice results from this paper and from Ref.~\cite{Kajantie:1995kf} are shown.
    \newtext{The critical region, where the discontinuity in the condensate obeys critical scaling and dives down to zero, appears to be very narrow on this logarithmic plot.}
    }
    \label{fig:phi2}
\end{figure}

\begin{figure}[t]
    \centering
    \includegraphics[width=0.48\textwidth]{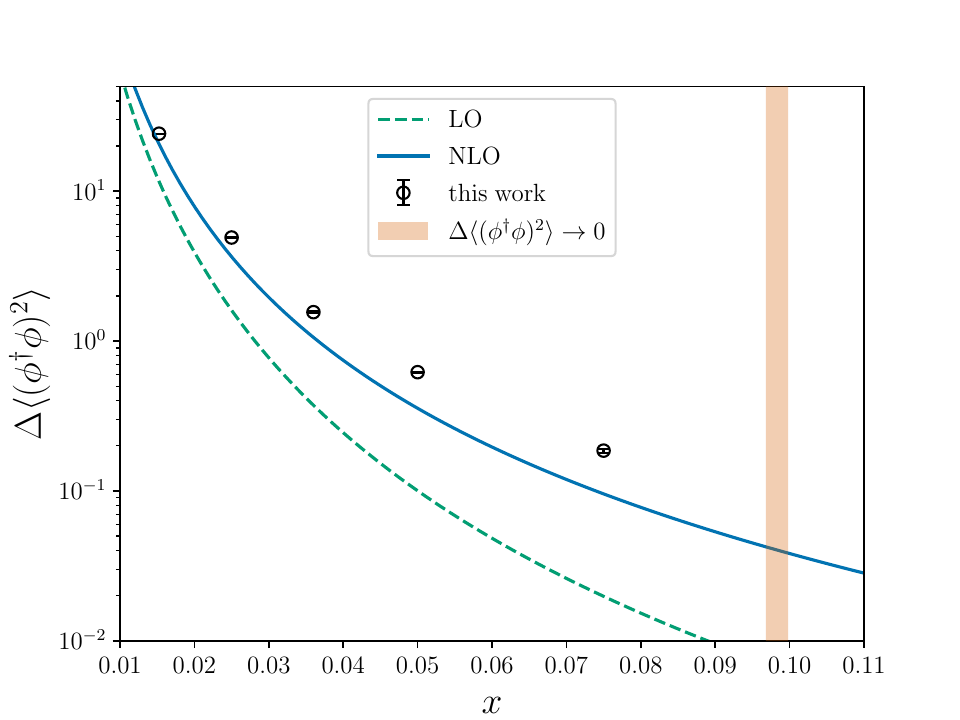}
    \caption{
    The discontinuity in the quartic Higgs condensate.
    This is the first time lattice results for this quantity have been reported.
    \newtext{Again the critical region appears to be rather narrow.}
    }
    \label{fig:phi4}
\end{figure}

Figs.~\ref{fig:phi2} and \ref{fig:phi4} show our continuum-extrapolated results for the quadratic and quartic Higgs condensates, together with existing results in the literature.
Extrapolations to infinite volume were carried out as for $\yc$.
For the extrapolations to zero lattice spacing, $a\to 0$, the linear term in the polynomial fit is absent for data with $O(a)$ improvement \cite{Moore:1996bf, Moore:1997np}.
To ensure the cancellation of $O(a)$ corrections, measurements of the condensates were carried out at $\yc(a)$, rather than at $\yc(a=0)$.
Some examples of our continuum extrapolations are collected in Appendix \ref{appendix:extrapolations}, demonstrating clearly the $O(a)$ improvement.
The complete data can be found at \cite{data}.

The LO and NLO perturbative expressions for the Higgs condensates are plotted alongside the lattice data in Figs.~\ref{fig:phi2} and \ref{fig:phi4}.
The expressions are given in Appendix \ref{appendix:perturbative_results}.
In both cases the NLO corrections are large, and move the perturbative results considerably closer to the lattice data.
Agreement between the lattice and perturbation theory is noticeably better for the quadratic condensate.
This is consistent with the perturbative expansion parameter, as given by the ratio NLO/LO, being smaller for the quadratic condensate.

\begin{figure}[t]
    \centering
    \includegraphics[width=0.48\textwidth]{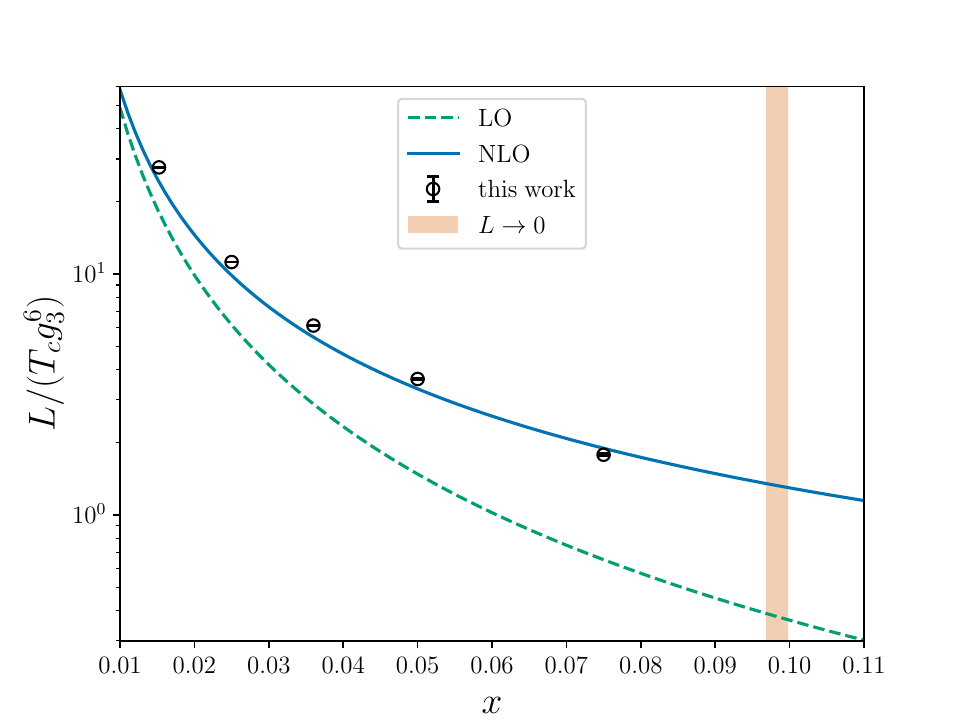}
    \caption{
    \newtext{The latent heat for an example using values $\eta_x(\Tc)=0.2$ and $\eta_y(\Tc)=4.6$ in Eq.~\eqref{eq:latent_heat_condensates}, illustrative of phase transitions in minimal extensions of the electroweak sector such as the xSM.}
    }
    \label{fig:latent_heat}
\end{figure}

\newtext{To compute the latent heat requires the coefficients of the Higgs condensates in Eq.~\eqref{eq:latent_heat_condensates}, $\eta_x$ and $\eta_y$ defined in Eq.~\eqref{eq:etas}. These are determined by modes with energies of order $O(\pi T)$ and above, and hence depend on the UV completion of the 3d EFT that we study. In the Standard Model, they take the values $\eta_x\sim -0.06$ and $\eta_y\sim 4.6$ in the vicinity of the electroweak crossover \cite{DOnofrio:2015gop}. In perturbative extensions of the Standard Model, the values are expected to be of the same order, as argued in Ref.~\cite{Gould:2019qek}. Let us consider the example of the xSM. Using the dimensional reduction relations of Ref.~\cite{Gorda:2018hvi}, and scanning over the one-step phase transitions considered in Ref.~\cite{Gould:2019qek}, one finds roughly that $|\eta_x(\Tc)|\lesssim 0.4$ and $|\eta_y(\Tc) - 4.6| \sim 0.6$. Fig.~\ref{fig:latent_heat} shows one such example with $\eta_x(\Tc)=0.2$ and $\eta_y(\Tc)=4.6$. In this case, for the strongest transition that we study at $x=0.0152473$, the quartic condensate gives a $\sim 20 \%$ contribution to the latent heat, decreasing for weaker transitions.}

\subsection{Surface tension} \label{sec:surface_tension}

The surface tension is the free energy per unit area of a macroscopically flat interface between phases, evaluated at the critical temperature.
It gives some measure of the strength of the transition, with the surface tension decreasing to zero as one approaches the endpoint of a line of first-order phase transitions, at which point the phases become miscible.
The surface tension is relevant to the bubble nucleation rate through the thin wall approximation, where the energy $E$ of a nucleating spherical bubble of radius $R$ is given by
\begin{align} \label{eq:thin_wall}
E = - \frac{4}{3}\pi R^3 p + 4\pi R^2 \sigma.
\end{align}
Here $p$ is the pressure, or free energy density difference, between the homogeneous phases.
The approximation is justified near the critical temperature, where the bubble radius is much larger than all other scales in the theory, in which case Eq.~\eqref{eq:thin_wall} receives corrections at $O(R)$.
Note that when working within the 3d EFT, it is more natural to work in terms of temperature-scaled quantities $S_3\equiv E/T$, $\Delta f_3 \equiv p/T$, and when working on the lattice we further scale out powers of $g_3^2$, so that everything computed is measured in units where $g_3^2=1$.
In this convention, the dimensionless surface tension is
\begin{align} \label{eq:sigma_T}
\sigma_3 \equiv \frac{\sigma}{T g_3^6}.
\end{align}

We measure the surface tension using the standard histogram method \cite{Binder:1982oqp}.
The known contributions of the capillary waves are accounted for in the infinite volume limit by fitting \cite{Bunk:1992pq, Iwasaki:1993qu, Moore:2000jw}
\begin{align}
\label{eq:surface_tension_volume}
\sigma_3 \cdot g_3^6 &= \frac{1}{2L_1^2}\log \frac{p_{\rm max}}{p_{\rm min}} \nonumber \\
&\quad + \frac{1}{2L_1^2}\left[\frac{3}{2}\log L_3 - \log L_1 + \text{const}\right].
\end{align}
Here $p_{\rm max}$ is the maximum of the order parameter histogram, equal to the probability density of being in either homogeneous phase,
and $p_{\rm min}$ is the minimum of the order parameter histogram between the two phases, equal to the probability density of being in a mixed phase containing two planar interfaces.
$L_1=L_2$ and $L_3$ are the physical lengths of the lattice in the 3 spatial directions.

\begin{figure}[t]
    \centering
    \includegraphics[width=0.48\textwidth]{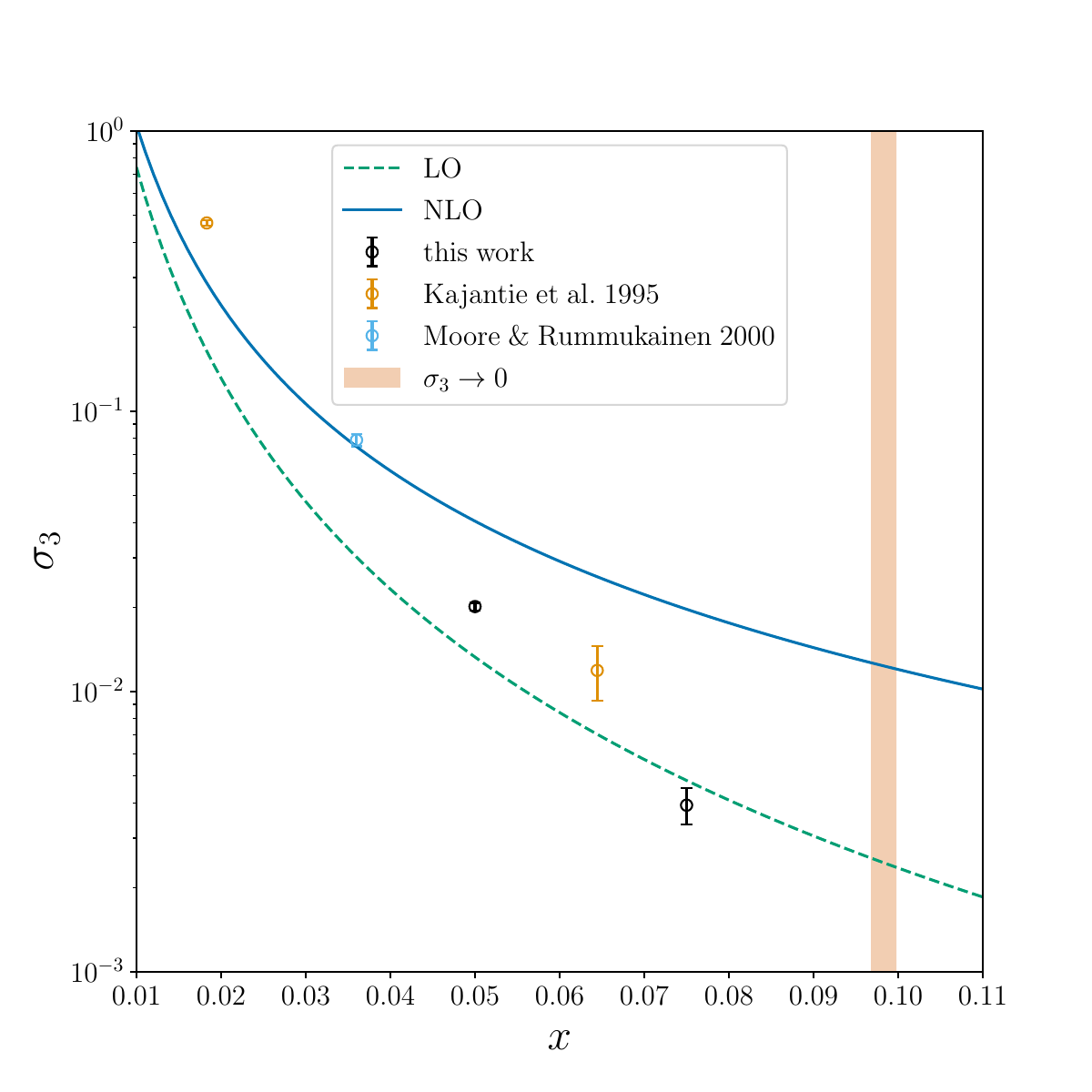}
    \caption{
    The surface tension plotted against $x$.
    Note that the continuum limit has not been taken for the lattice point at smallest $x$.
    Lattice results from this paper and from Refs.~\cite{Kajantie:1995kf, Moore:2000jw} are shown.
    \newtext{The approach to zero in the critical region is clearer at smaller $x$ for the surface tension than for the latent heat. Small values of the surface tension, while the latent heat is still sizeable, has also been observed in the 2HDM \cite{Kainulainen:2019kyp}.}
    }
    \label{fig:surface_tension}
\end{figure}

An alternative method to calculate the surface tension, by Fourier analysing the spectrum of transverse fluctuations of a phase interface, was presented in Ref.~\cite{Moore:1996bn}, and also used in Ref.~\cite{Moore:2000jw}.
This method has been shown to give more accurate results with less computational resources.
However, as the computation of the surface tension is not a major goal of our paper, we have not pursued this method.

For measurements of the surface tension, typically cylindrical lattices with one size much longer than the others, $L_3\gg L_1, L_2$, are used so that interactions between interfaces are small.
We have not carried out any dedicated simulations on such long lattices, but have merely attempted to reuse the results from the simulations for Secs.~\ref{sec:critical_temperature} and \ref{sec:latent_heat} on cubic lattices.
As a consequence, many of our simulations are far from the infinite volume limit, and the volume dependence of the surface tension showed marked deviations from the expected form \eqref{eq:surface_tension_volume}.
This was particularly the case for the stronger transitions at smaller $x$, so we have only reported results for $x\geq 0.05$, where the infinite volume extrapolations seemed under better control.
\newtext{Some example plots of the infinite volume extrapolations are given in Appendix \ref{appendix:extrapolations}.}
Extrapolations to zero lattice spacing were performed as in Sec.~\ref{sec:latent_heat}.

Our continuum extrapolated results are shown in Fig.~\ref{fig:surface_tension}, together with existing results from the literature.
We also include the LO and NLO perturbative approximations to the surface tension, for which the expressions can be found in Appendix \ref{appendix:perturbative_results}.
Fig.~\ref{fig:surface_tension} shows that there is relatively poor agreement between the lattice and perturbative results for the surface tension.

\subsection{Summary of equilibrium results} \label{sec:equilibrium_summary}

Our equilibrium lattice results are collected in Table \ref{table:results_yc}.
We have also included lattice results from the literature in which the continuum limit was taken \cite{Kajantie:1995kf, Rummukainen:1998as, Laine:1998jb, Moore:2000jw}.
For completeness, we note that there are a number of other lattice results for this model in the literature at finite lattice spacing \cite{Farakos:1994kj, Ilgenfritz:1995sh, Gurtler:1995ix, Gurtler:1996wx, Karsch:1996yh, Gurtler:1997nm, Gurtler:1997hr}.

\begin{table}[t]
	\centering
    \begin{tabular}{lllll}
      \hline
      $x$ & $\yc$ & $\Delta \langle \phi^\dagger\phi \rangle_{\rm c}$ & $\Delta \langle (\phi^\dagger\phi)^2 \rangle_{\rm c}$ & $\sigma_3$ \\
      \hline
      0.0152473 & 0.071998(79) & 4.9635(54) & 24.184(44) & \multicolumn{1}{c}{---} \\
	  0.0183 & 0.05904(56) & 4.07(13) & \multicolumn{1}{c}{---} & [0.47(1)] \\ 
      0.025 & 0.04015(10) & 2.2207(81) & 4.906(64) & \multicolumn{1}{c}{---} \\ 
      0.036 & 0.02058(29) & 1.2588(86) & 1.558(21) & 0.079(4) \\
      0.05 & 0.00692(11) & 0.7692(44) & 0.6198(52) & 0.02011(62) \\ 
      0.06444 & -0.00146(35) & 0.491(8) & \multicolumn{1}{c}{---} & 0.0119(26) \\ 
      0.075 & -0.00827(20) & 0.3780(47) & 0.1857(59) & 0.00393(58) \\ 
      0.08970 & -0.01531(69) & 0.302(18) & \multicolumn{1}{c}{---} & \multicolumn{1}{c}{---} \\
      \hline
    \end{tabular}
  \caption{
  Lattice Monte-Carlo results for equilibrium quantities describing first-order electroweak phase transitions, from this work, and from Refs.~\cite{Kajantie:1995kf,Moore:2000jw}.
  The result in square brackets has not been extrapolated to the continuum limit; it is for a single, nonzero lattice spacing.
  \label{table:results_yc}}
\end{table}

Perturbation theory demonstrates relatively good agreement with the lattice for sufficiently small $x$, and diverges from it at larger $x$.
However, the precise level of agreement depends on the observable, with the surface tension showing the greatest discrepancies between lattice and perturbation theory.

\section{Bubble nucleation} \label{sec:nucleation}

Bubble nucleation is the first stage of the dynamical evolution of a first-order phase transition. After nucleation, the pressure difference between phases causes the bubbles to grow acceleratedly, until it reaches a terminal velocity where the friction of the plasma balances against the bubbles' outward pressure. The growing bubbles eventually meet, colliding and creating sound waves which in turn create gravitational waves \cite{Hindmarsh:2013xza, Hindmarsh:2015qta, Hindmarsh:2017gnf}. The average distance between bubbles, and consequently the peak in the gravitational wave spectrum, is determined by the bubble nucleation rate.
Therefore, developing a quantitative description of bubble nucleation is crucial for relating gravitational wave observations to the physics of cosmological phase transitions.

To calculate the bubble nucleation rate on the lattice, we make use of the \newtext{Langevin description of the dynamics of the infrared modes of non-Abelian gauge theories at} high-temperature \cite{Bodeker:1998hm, Bodeker:1999ey, Bodeker:1999ud, Arnold:1998cy, Arnold:1999jf, Arnold:1999uy, Litim:1999ns, Litim:1999id, Litim:1999ca, Moore:1998zk, Moore:2000mx}. \newtext{For the SU(2) Higgs theory, the relevant Langevin equations read \cite{Moore:2000mx, Moore:2000jw}}
\newtext{%
\begin{align} \label{eq:langevin_gauge}
\sigma_\text{el} (D_t A_i)^a &= - \frac{\delta H}{\delta A_i^a} + \xi_i^a, \\
\sigma_\text{el} D_t \phi &= -\eta \frac{\delta H}{\delta \phi^\dagger} + \xi_\phi \label{eq:langevin_higgs}
\end{align}
where $H=\int_\mb{x} \ms{L}_3$, and $\ms{L}_3$ is the Euclidean Lagrangian of the 3d EFT, Eq.~\eqref{eq:lagrangian}. The parameter $\sigma_\text{el}\sim T/\log 1/g$ is the SU(2) colour conductivity, and the noise terms $\xi_i^a$ and $\xi_\phi$ satisfy
\begin{align}
\langle \xi_i^a(t, \mb{x}) \xi_j^b(u, \mb{y}) \rangle &=
2 \sigma_\text{el} \delta_{ij}\delta^{ab}
\delta(\mb{x} - \mb{y})\delta(t - u), \\
\langle \xi_\phi(t, \mb{x}) \xi_\phi^\dagger (u, \mb{y}) \rangle &=
2 \eta \sigma_\text{el} \mathds{1}
\delta(\mb{x} - \mb{y})\delta(t - u),
\end{align}%
where $\eta \sim 1/g^2 \gg 1$ is the ratio of evolution rates for the Higgs field to the gauge bosons, and $\mathds{1}$ is the unit matrix for the fundamental SU(2) indices.}

For the non-perturbative \newtext{infrared} gauge fields of the symmetric phase this describes the full quantum dynamics, up to corrections of order $O(1/\log 1/g)$.
We will assume the description also applies in the vicinity of the critical bubble, though the original derivations of the Langevin equations did not consider this case.
The hard-thermal loop effective theory offers a more accurate dynamical description \cite{Braaten:1989mz, Frenkel:1989br, Taylor:1990ia, Frenkel:1991ts, Braaten:1991gm, Blaizot:1993zk, Blaizot:1993be, Nair:1993rx}, which is correct up to $O(g)$ corrections, and for which numerical schemes exist \cite{Hu:1996sf, Moore:1997sn, Bodeker:1995pp, Bodeker:1999gx, Iancu:1998bmf}. \newtext{However, hard-thermal loop effective theory approaches do not have a continuum limit, and thus we do not pursue them here.}
The nucleation rate is of cosmological relevance when it is roughly $e^{-100}$ \cite{Enqvist:1991xw, Anderson:1991zb}, in which case an error of $\sim e^{\pm 1}$ is only a 1\% error on the logarithm of the rate.

In principle, one could calculate the bubble nucleation rate by simply initialising a lattice in the metastable phase, evolving it according to the Langevin equations and waiting to see how long it takes to decay into the stable phase \cite{PhysRevB.42.6614, Alford:1991qg, Alford:1993ph, Alford:1993zf, Borsanyi:2000ua}.
However, bubble nucleation is an exponentially suppressed process, and for the cosmologically relevant situation the exponent is $O(100)$.
Thus, unfeasibly long simulations times would be required to simulate bubble nucleation directly.

Instead, we adopt the method proposed in Refs.~\cite{Moore:2000jw, Moore:2001vf}.
A variant of this method was also used to compute the broken-phase sphaleron rate \cite{Moore:1998swa, Moore:1998ge}.
The approach follows the spirit of Langer's seminal work \cite{Langer:1969bc}, and generalises it beyond the saddlepoint approximation.

When bubble nucleation is very slow, the metastable phase will have time to equilibrate long before nucleation occurs. The relative probability of a field configuration on the metastable side, including bubble configurations, is then given by its Boltzmann weight. This {\em statistical} problem, of finding the probability of bubble configurations, can be calculated with standard Monte-Carlo methods. To turn the resulting probability into a rate, some additional {\em dynamical} information is required, about how often the dynamical end state of a bubble configuration is indeed in the stable phase, and about the rate of bubble growth. Calculating these dynamical quantities only requires relatively short Langevin simulations. In this way, the difficult problem of simulating an exponentially suppressed process is factorised into two easier problems: one statistical and the other dynamical.

\subsection{Nucleation rate} \label{sec:nucleation_rate}

To compute the nucleation rate, we follow closely the approach of Refs.~\cite{Moore:2000jw, Moore:2001vf}, where the reader can find further details.
This approach requires an observable which is phase-sensitive, which we take to be the spatially averaged Higgs quadratic condensate, $\phi^\dagger\phi$, as this allows us to reweight in $y$.
The probability distribution $p(\phi^\dagger\phi)$ shows a two-peak structure, with the peak at smaller (larger) values of $\phi^\dagger\phi$ corresponding to the symmetric (broken) phase; see Fig.~\ref{fig:hists}.
Any classical real-time trajectory going between symmetric and broken phases must pass through intermediate values of $\phi^\dagger\phi$, which are exponentially unlikely in the equilibrium distribution.
On an appropriately sized lattice,%
\footnote{\newtext{As demonstrated in Refs.~\cite{Moore:2000jw, Moore:2001vf}, an appropriately sized lattice must be sufficiently large that the bubble takes up a volume fraction less than about $4\pi/81\approx 15\%$. On the other hand, in order that only one bubble is accommodated, the lattice extent should be much smaller than the average distance between nucleating bubbles in infinite volume. In practice this latter condition is irrelevant, as the exponential suppression of bubble configurations implies that their average separation is exponentially large.}
}
the minimum of $p(\phi^\dagger\phi)$ between phases consists of spatially localised {\em critical bubbles}, the gatekeepers between phases.

\begin{figure}[t]
    \centering
    \includegraphics[width=0.48\textwidth]{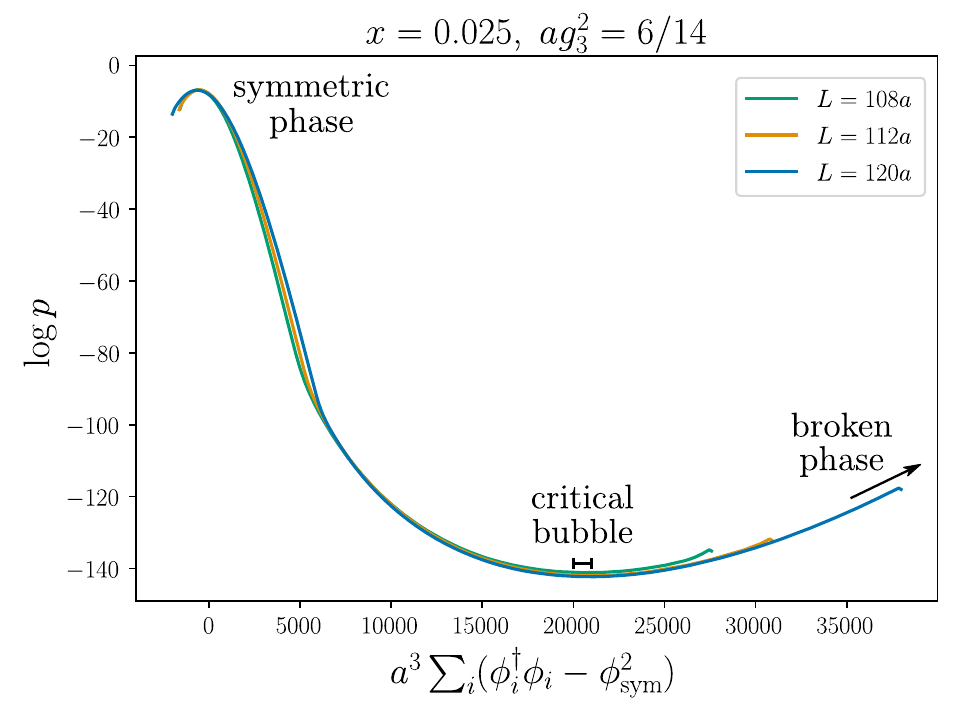}
    \caption{
    Histograms showing the probability distribution of the volume-integrated quadratic Higgs condensate, normalised with respect to its value in the symmetric phase.
    This observable has been plotted, instead of the volume-averaged quadratic Higgs condensate, because the critical bubble has fixed absolute volume, and hence its position on this plot agrees for different lattice sizes.
    The relative probability of the critical bubble versus the symmetric phase is what determines the statistical part of the nucleation rate.
    This quantity is stable under changes of the lattice size, $L$, provided the bubble takes up a volume fraction $\lesssim 4\pi/81$ \cite{Moore:2000jw}.
    }
    \label{fig:hists_nucleation}
\end{figure}

The process of nucleation is limited by the probability of critical bubbles.
We denote by $\phi^\dagger\phi=\varphi^2_\text{C}$ the location of the minimum of $p(\phi^\dagger\phi)$.
This condition defines a co-dimension one surface in configuration space, separating the two phases, called the critical surface or separatrix.
The probability $P$ of being in some small vicinity $\epsilon$ around the critical surface is
\begin{align}
P\left(|\phi^\dagger\phi-\varphi^2_\text{C}| < \epsilon/2\right) &= 
\int_{\varphi^2_\text{C}-\epsilon/2}^{\varphi^2_\text{C}+\epsilon/2} p(\phi^\dagger\phi) d(\phi^\dagger\phi) , \\
&\approx p(\varphi^2_\text{C}) \epsilon.
\end{align}
In the context of the nucleation rate, this should be normalised relative to the probability of the metastable phase $P\left(\phi^\dagger\phi<\varphi^2_\text{C}\right)$.
Some examples of histograms are shown in Fig.~\ref{fig:hists_nucleation}, where the region $|\phi^\dagger\phi-\varphi^2_\text{C}| < \epsilon/2$ has been highlighted and labelled ``critical bubble''.

To attain the probability flux through the critical surface in configuration space, one needs to multiply the probability density at the critical bubble by the magnitude of the vector perpendicular to the critical surface, $\left| \Delta (\phi^\dagger\phi)/\Delta t \right|_{\varphi^2_\text{C}}$.

\begin{figure}[t]
    \centering
    \includegraphics[width=0.48\textwidth]{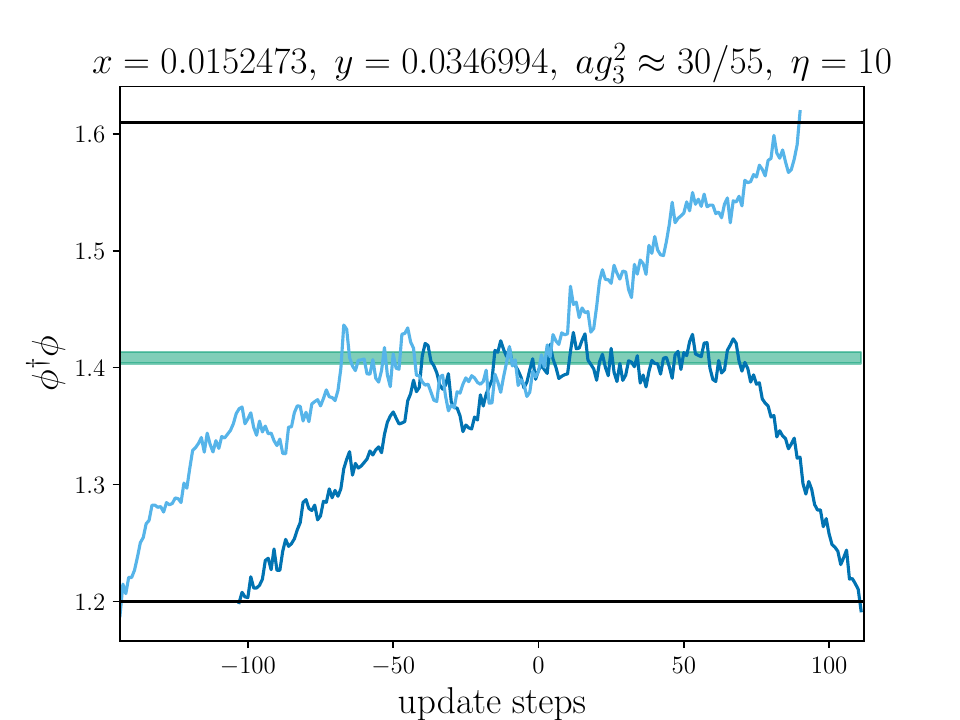}
    \caption{Time evolution of the scalar condensate in the vicinity of the critical bubble. The light and dark blue lines show two different instances of the time evolution starting from the same initial bubble configuration at update step 0. The shaded green band shows the region satisfying $|\phi^\dagger\phi-\varphi^2_\text{C}| < \epsilon/2$, and the horizontal black lines show where the simulation is ended as the bubble grows/shrinks towards the homogeneous phases. A 3d video of one such trajectory can be seen at Ref.~\cite{videos}.}
    \label{fig:trajectories}
\end{figure}

Finally, to attain the nucleation rate from the probability flux, we must account for the fact that trajectories can cross the critical surface more than once, and may in fact cross an even number of times and hence not tunnel.
Some example trajectories are shown in Fig.~\ref{fig:trajectories}.
To account for this effect, one introduces the following dynamical prefactor,
\begin{align}
\mb{d}
&=
\frac{\delta_\text{tunnel}}{N_\text{crossings}} ,
\end{align}
where $\delta_\text{tunnel}=1$ if there is tunnelling and $\delta_\text{tunnel}=0$ if not, and $N_\text{crossings}$ is the number of crossings of the separatrix.
From the definition, one can see that $\mb{d}$ lies between zero and one.
If all critical bubbles were to be equally likely to expand to the broken phase or contract to the symmetric phase, and all were to only cross the critical surface either once or twice, then $\langle\mb{d}\rangle=1/2$. In fact, one expects $\langle\mb{d}\rangle<1/2$ due to the spiky nature of Langevin evolution causing multiple crossings; see Fig.~\ref{fig:trajectories}.
Conversely $\mb{d}$ increases as the trajectory is downsampled, though this is compensated for by a decrease in the magnitude of $\left| \Delta (\phi^\dagger\phi)/\Delta t \right|_{\varphi^2_\text{C}}$, such that the product is independent of the sampling rate \cite{Moore:2000jw}.

\begin{figure}[t]
    \centering
    \includegraphics[width=0.48\textwidth]{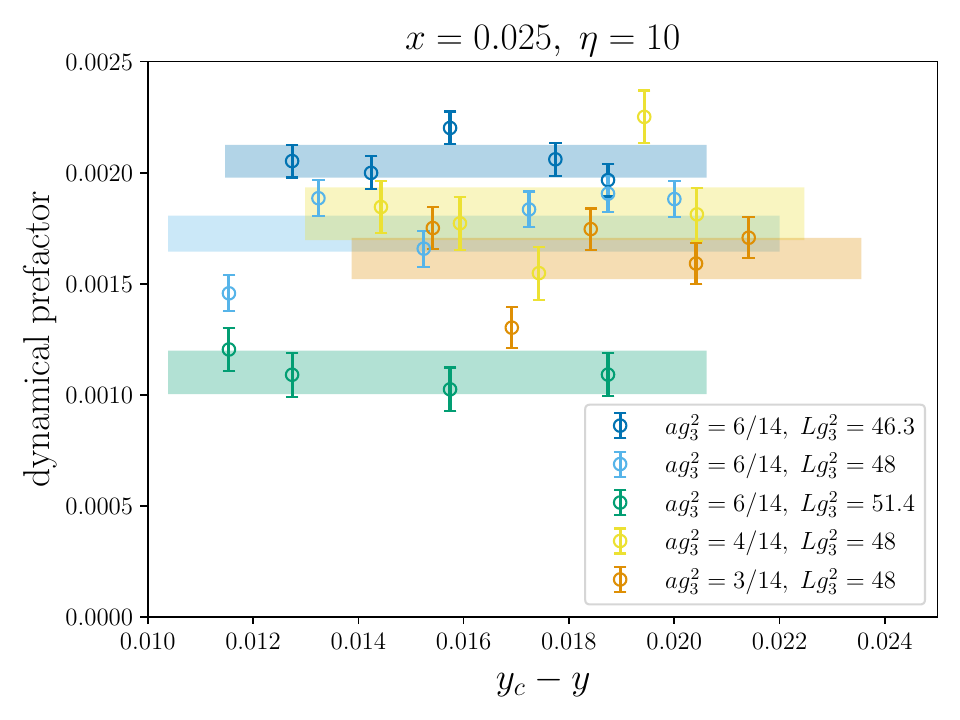}
    \caption{The dynamical prefactor of the nucleation rate evaluated on a range of lattices, and plotted as a function of the degree of supercooling, $\yc-y$.
    For each given lattice spacing and volume, fits to a constant independent of $\yc-y$ are shown as the coloured bands.}
    \label{fig:prefactor}
\end{figure}

In sum, the nucleation rate per unit volume, as calculated on the lattice, is given by the following expression \cite{Moore:2000jw, Moore:2001vf}
\begin{align} \label{eq:rate_factors}
\rate =
\underbrace{
\left\langle
\frac{\mb{d}}{2}
\left| \frac{\Delta (\phi^\dagger\phi)}{\Delta t}\right|_{\varphi^2_\text{C}}
\right\rangle
}_\text{dynamical}
\underbrace{
\frac{P\left(|\phi^\dagger\phi-\varphi^2_\text{C}| < \epsilon/2\right)}{\epsilon V  P\left(\phi^\dagger\phi<\varphi^2_\text{C}\right)}
}_\text{statistical}.
\end{align}
The statistical part of this expression may be calculated with standard (multicanonical) Monte-Carlo simulations.
The dynamical part requires real-time Langevin simulations.
The factor of 1/2 removes overcounting, because half of the transitions between the symmetric and broken phase are in the wrong direction.

Compared to a simple wait-and-see approach \cite{PhysRevB.42.6614, Alford:1991qg, Alford:1993ph, Alford:1993zf, Borsanyi:2000ua}, Eq.~\eqref{eq:rate_factors} relies on the crucial assumption that the full rate factorises into dynamical and statistical parts.
This can be argued for based on the enormous hierarchy of scales between the total timescale of nucleation, and the timescale for which the system remains in the vicinity of the critical bubble.
Over the exponentially long times taken to get from the symmetric phase to the vicinity of the critical bubble, all correlations in time are expected to be washed out by the large numbers of interactions which occur, so that this process occurs effectively in equilibrium.
Only for the short time spent in the vicinity of the critical bubble is it necessary to account for correlations in time.
Such a factorisation can be explicitly demonstrated within a saddlepoint approximation of the path integral \cite{Langer:1969bc, Gould:2021ccf, Ekstedt:2022tqk}.
However, while the present approach relies on this factorisation, lattice simulations go beyond the saddlepoint approximation.

\newtext{To numerically simulate the Langevin equations, Eqs.~\eqref{eq:langevin_gauge} and \eqref{eq:langevin_higgs}, one could directly discretise the time derivative, and evolve as a stochastic initial value problem.
However, in this approach finite time step errors will lead to deviations from the correct thermodynamics of the system, which may cause havoc for the evolution of a finely balanced critical bubble.
Alternatively, one may perform simulations using a different dissipative update, if the relationship to Langevin evolution is known.
Following Ref.~\cite{Moore:2000jw}, we choose to perform}
heatbath updates on the gauge fields, which is equivalent to Langevin evolution \newtext{where the Langevin time step $\Delta t$ is related to the number of heatbath updates per link $n_\text{hb}$ through
\begin{align}
\Delta t = \frac{1}{4} n_\text{hb} \sigma_\text{el} a^2.
\end{align}}
\newtext{This relation was proven for the case where} the lattice sites are updated in a random order \cite{Moore:1998zk}.
We however adopt a checkerboard order, which is significantly simpler to parallelise.
This can be partially justified by noting that for each group of odd or even sites updates are uncorrelated, therefore a sequentially ordered update algorithm is equivalent to a random ordered algorithm within each group.
In addition, we randomly select whether odd or even sites are updated in each sweep of the lattice.

\begin{figure}[t]
    \centering
    \includegraphics[width=0.48\textwidth]{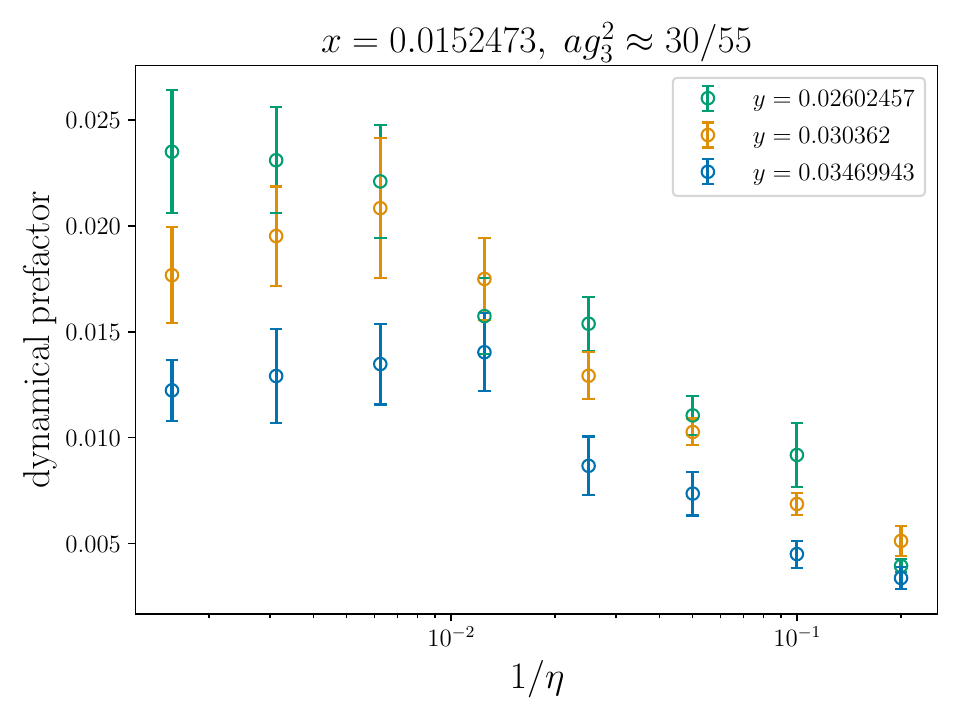}
    \caption{The dependence of the dynamical prefactor on $\eta$, the number of Higgs field updates for each gauge field update.
    The limit $1/\eta \to 0$ gives the leading order result.}
    \label{fig:prefactor_eta}
\end{figure}

The evolution of the Higgs field is parametrically faster than that of the gauge fields, therefore, to leading order in the couplings, the gauge field sees the Higgs field in equilibrium \cite{Moore:1998zk, Moore:2000jw}.
To ensure this, for every one gauge-field update, we perform a large number $\eta \gg 1$ of heatbath and overrelaxation updates on the Higgs field.
Fig.~\ref{fig:prefactor_eta} shows how the dynamical prefactor depends on $\eta$ at $x=0.0152473$.
This shows that the prefactor initially grows with $\eta$ until eventually flattening off, suggesting a smooth $\eta \to \infty$ limit.
The result of taking this limit gives the leading order result \cite{Moore:1998zk}.
We have simply used either $\eta=10$ or $\eta=40$ in our final results (for specifics see \cite{data}).
The error thereby introduced is comparable to the uncertainty in the statistical part of the rate.

The $y$ dependence of the statistical part of the nucleation rate can be determined by reweighting, following Eq.~\eqref{eq:reweighting}.
This greatly reduces the computational effort of the calculation, as for a given value of $x$, the whole functional dependence on $y$ follows from one simulation at a single value of $y$.
There is unfortunately no such trick applicable to the dynamical part of the nucleation rate.
We have thus studied the $y$ dependence of the dynamical part through direct simulations, some examples of which are shown in Fig.~\ref{fig:prefactor}.
As can be seen, the $y$ dependence is very mild, being consistent with constant within errors.

Fig.~\ref{fig:prefactor} also reveals some volume dependence of the dynamical part.
However, when incorporated into the full nucleation rate the total volume dependence is milder, consistent with volume-independent within errors.
The finite size of $\epsilon$ may be responsible for the volume dependence of the dynamical part.
This is because, as the volume grows the minimum of the probability distribution for $\phi^\dagger\phi$ becomes more strongly curved, leading to bubbles at the edges of the range $|\phi^\dagger\phi-\varphi^2_\text{C}| < \epsilon/2$ being more represented in the dynamical prefactor calculation.
These bubbles at the edges of the range are less likely to tunnel, and hence for them $\mb{d}$ is smaller.
While this effect decreases the dynamical part, it increases the statistical part, because the integrated probability $P\left(|\phi^\dagger\phi-\varphi^2_\text{C}| < \epsilon/2\right)$ grows due to the increased probability density at the edges of the range.
These two effects cancel, as argued in the appendix of Ref.~\cite{Moore:2000jw}.

The relationship between the physical nucleation rate $\Gamma$, and the rescaled quantity calculated on the lattice, $\rate$, takes the form
\begin{align} \label{eq:rate_relation}
\Gamma = \frac{g_3^{4}}{\sigma_\text{el}}\cdot (g_3^{2})^3 \cdot \rate(x, y).
\end{align}
The prefactors on the right hand side of Eq.~\eqref{eq:rate_relation} arise from the relation between physical units and those used on the lattice \cite{Moore:1998zk}.
The factors also reflect properties of the typical infrared gauge configurations relevant to bubble nucleation, with the first factor being the inverse time scale of their evolution, and the second factor being their inverse volume scale.

In a sufficiently large lattice, finite volume effects on bubble nucleation are exponentially small, owing to the model being gapped.
This requires that the bubble fits inside the lattice with enough space around that it does not interact with itself through the periodic boundary conditions.
As shown in Refs.~\cite{Moore:2000jw, Moore:2001vf}, this is equivalent to the condition the bubble takes up less than an approximately $\pi^2/81$ fraction of the total lattice volume.
In smaller lattices, the lowest energy configuration is distorted away from spherical, becoming either cylindrical or slab-like.
Transitions between these discretely different geometrical configurations appear as kinks in the histograms.
There is another kink for the transition between bulk fluctuations and localised bubble fluctuations \cite{Moore:2000jw, Moore:2001vf, biskup2002formation, binder2003theory, Nu_baumer_2006}.

\begin{figure}[t]
    \centering
    \includegraphics[width=0.48\textwidth]{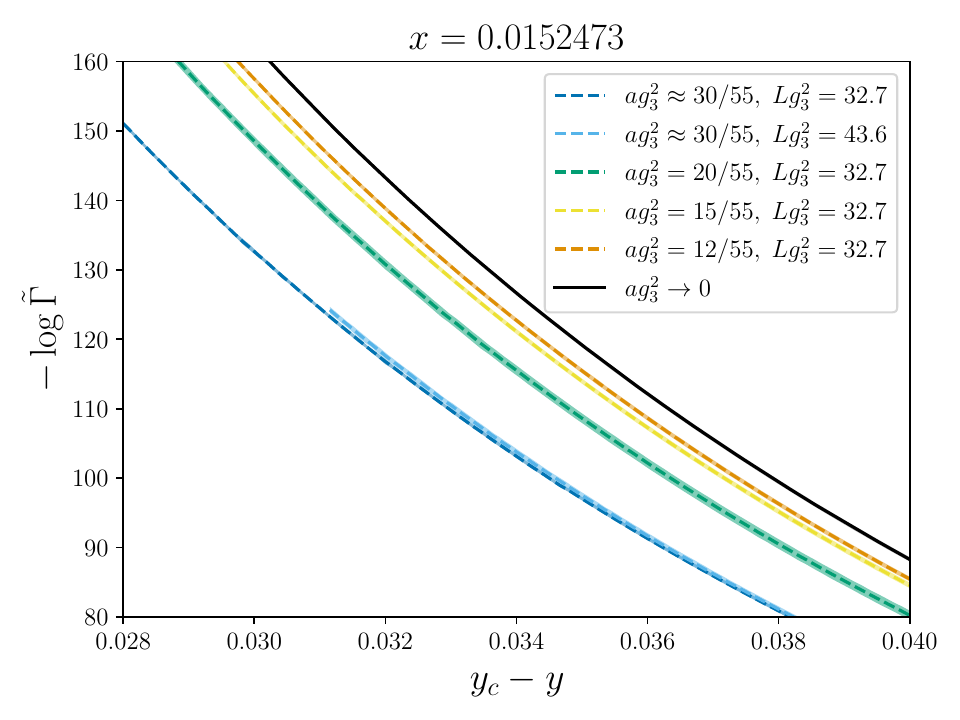}
    \includegraphics[width=0.48\textwidth]{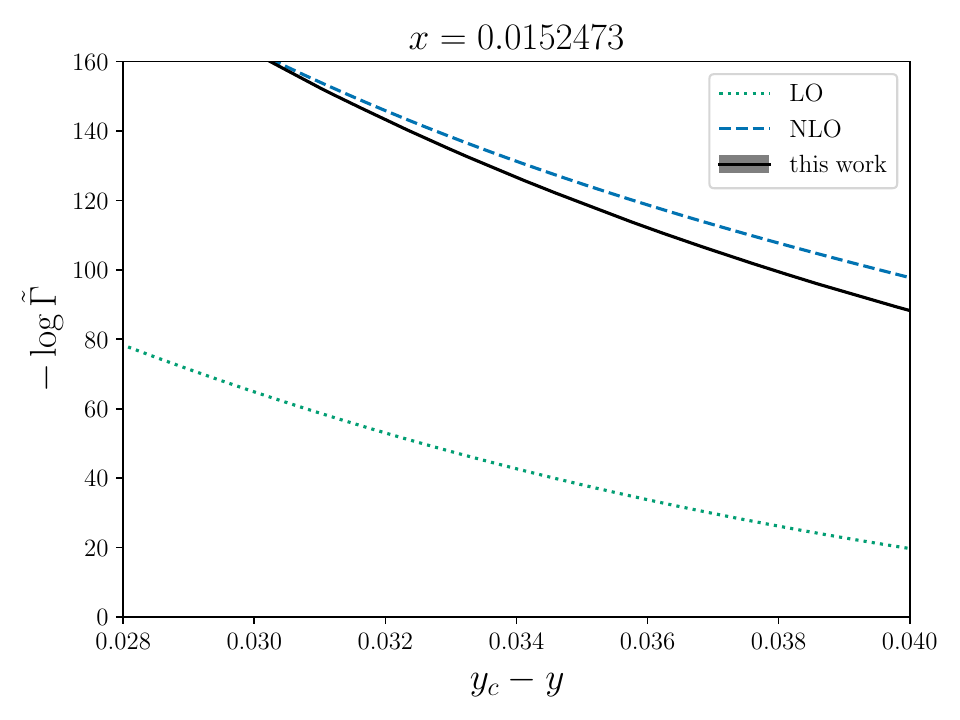}
    \caption{
    The bubble nucleation rate at the strongest transition which we have studied.
    Above we show data on finite lattices together with the continuum extrapolation.
    Below the extrapolated rate is compared to two perturbative approximations.
    }
    \label{fig:rate}
\end{figure}

\begin{figure}[t]
    \centering
    \includegraphics[width=0.48\textwidth]{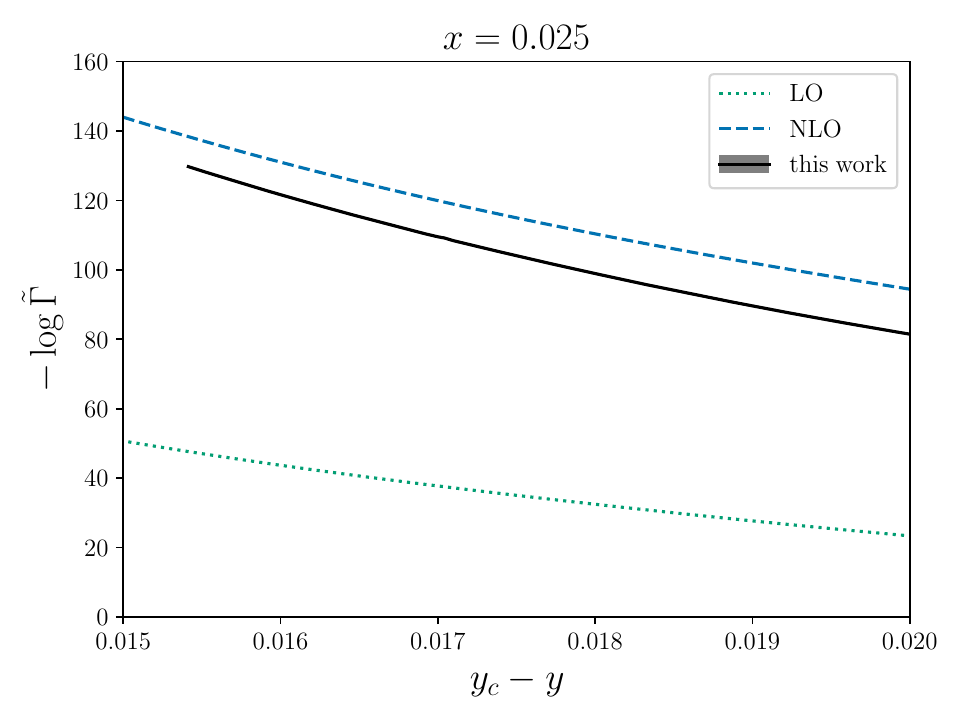}
    \includegraphics[width=0.48\textwidth]{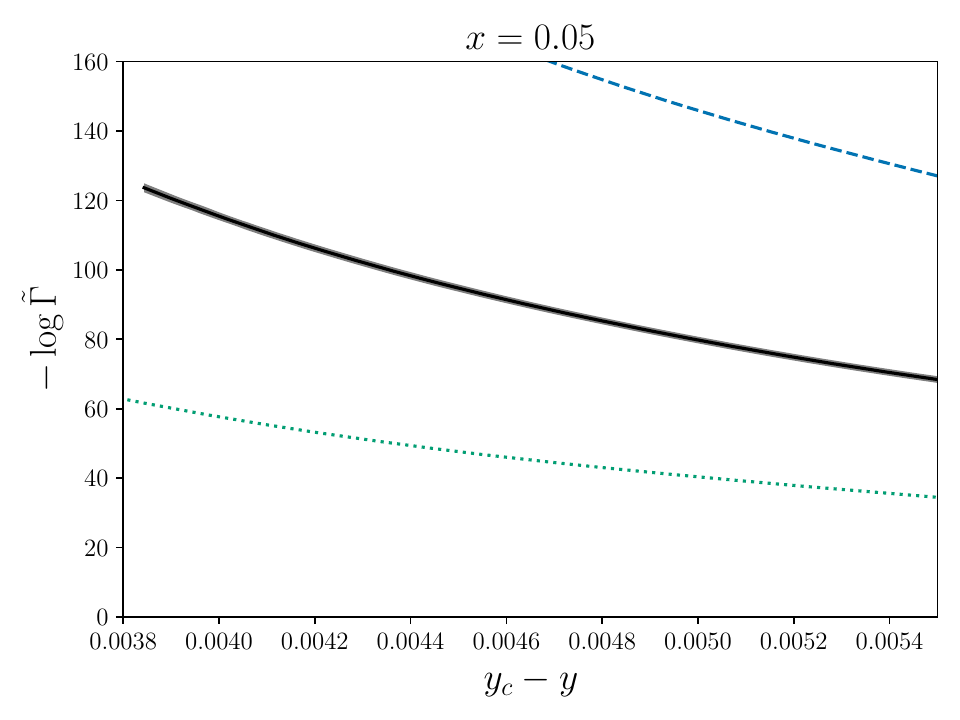}
    \caption{
    The bubble nucleation rate at two weaker transitions which we have studied, together with two perturbative approximations.
    }
    \label{fig:rates}
\end{figure}

Fig.~\ref{fig:rate} shows our results for the nucleation rate $x=0.0152473$ on a range of lattices, together with the continuum extrapolation.
Finite lattice results and continuum extrapolations for the other two values of $x$ simulated, 0.025 and 0.05, are presented in Appendix \ref{appendix:extrapolations}.
All lattices were chosen to be sufficiently large to accommodate spherical bubbles.
For the smallest lattice spacings, a number of different lattice volumes are shown, the results of which agree within error.
This demonstrates that our lattices are sufficiently large that we may neglect volume dependence.
\newtext{Table~\ref{table:nucleation_lattices} lists the complete set of parameters and lattice volumes on which we have computed the nucleation rate.}

\begin{table}[t]
	\centering
    \begin{tabular}{llc}
      \hline
      \muco{$x$} & \muco{$a g_3^2$} & \muco{volumes$/a^3$} \\
      \hline
      0.0152473 & 0.5455087 & $60^3$, $80^3$ \\
      ~ & 0.3636364  & $90^3$ \\
      ~ & 0.2727273  & $120^3$ \\
      ~ & 0.2181818  & $150^3$ \\
      0.025 & 0.4285714  & $108^3$, $112^3$, $120^3$ \\
      ~ & 0.2857143  & $168^3$ \\
      ~ & 0.2142857  & $224^3$ \\
      0.05 & 0.5  & $168^3$ \\
      ~ & 0.3333333  & $224^3$ \\
      \hline
      \end{tabular}
  \caption{\newtext{Lattices used for the simulations of bubble nucleation. A complete list of runs can be found at \cite{data}. Note that  larger lattices were required at larger $x$, because, due to the small surface tension, the cosmologically relevant critical bubbles, for which $-\log\rate\sim 100$, are closer to the thin-wall limit. All volumes are sufficiently large that the bubble takes up a small volume fraction $(\varphi_\text{C}^2 - \varphi_\text{S}^2)/(\varphi_\text{B}^2 - \varphi_\text{S}^2)\lesssim 4\pi/81$, ensuring that the bubble does not see itself \cite{Moore:2000jw}. Here $\varphi_\text{S}^2$ and $\varphi_\text{B}^2$ refer to the positions of the peaks of the symmetric and broken phases respectively.}
  \label{table:nucleation_lattices}}
\end{table}

Also shown in Fig.~\ref{fig:rate} are the effects of changing the lattice spacing, while keeping the physical volume fixed.
We perform continuum extrapolations by fitting $1+a^2$, for each value of $\yc(a)-y$.
This is justified because all our simulations were performed with an $O(a)$ improved lattice action.
The only place where $O(a)$ corrections may in principle arise is through the relation between Langevin and heatbath updates \cite{Moore:1998zk}, though we expect such corrections to be small, if present, as they only affect the dynamical prefactor.

The calculated nucleation rate is a function of $y$, with error bars, for each value of $x$.
For convenience of future use, we fit our numerical results to the following function
\begin{align}
\action(x,y) = \frac{s_{-2}}{(\yc-y)^2} + \frac{s_{-1}}{\yc-y} + s_0.
\end{align}
For each value of $x$ we perform the fit over the range of $\yc-y$ shown in Figs.~\ref{fig:rate} and \ref{fig:rates}.
This function is motivated by the expected form in the thin-wall limit, $(\yc-y)\to 0_+$.
In this limit, the action grows as
\begin{align}
\action(x,y) \to \frac{16\pi\sigma_3^3}{3(\Delta\langle\phi^\dagger\phi\rangle_\text{c})^2 (\yc-y)^2},
\end{align}
here written in terms of the surface tension and jump in the quadratic condensate, each evaluated at $y=\yc$.
Note, however, that we do not perform the fit in the limit $y\to \yc$, but rather in some range of nonzero $\yc-y$.
Thus, our fit results $s_i$ are {\em not} the coefficients in an expansion about $y=\yc$, but simply parameterise our results over the range of $y$ studied.
The results are collected in Table \ref{table:results_rate}.

\begin{table}[t]
	\centering
    \begin{tabular}{llll}
      \hline
      \muco{$x_\text{c}$} & \muco{$s_{-2}$} & \muco{$s_{-1}$} & \muco{$s_{0}$} \\
      \hline
      0.0152473 & 0.0506(66) & 5.93(40) & -91.9(6.0) \\
      0.025 & 0.01419(48) & 1.612(56) & -35.0(1.6) \\
      0.036$^*$ & 0.0061(30) & 0.15(64) & -1(34) \\
      0.05$^*$ & 0.001410(99) & 0.081(40) & 7.2(4.5) \\
      \hline
    \end{tabular}
  \caption{Lattice Monte-Carlo results for fits to the bubble nucleation rate, from this work, and from Ref.~\cite{Moore:2000jw}.
  The (large) errors for $x=0.036$ follow from assuming $\pm 1$ errors on the rate \cite{Moore:2000jw}.
  The asterisks $(^*)$ are a reminder that, for these values of $x$, only two lattice spacings have been used to extrapolate to $a\to 0$.
  \label{table:results_rate}}
\end{table}

Constructing the perturbative expansion for the nucleation rate requires some care.
We utilise the EFT approach to bubble nucleation \cite{Gould:2021ccf}, and perform a strict expansion in $x$ to ensure order-by-order gauge and renormalisation-scale invariance \cite{Gould:2021ccf, Hirvonen:2021zej, Lofgren:2021ogg}.
Details are given in Appendix \ref{appendix:perturbative_results}.
Beyond LO, the difference between the values of $\yc$ in successive approximations poses difficulties for a strict expansion, because at $\yc$ the rate is nonanalytic and the logarithm of the rate is singular.
This issue can be overcome by computing the rate at fixed $\delta y \equiv \yc - y$, rather than at fixed $y$.
At NLO, this means that one writes $y = \yc^\text{LO} + \yc^\text{NLO} - \delta y$, and then splits the LO and NLO parts of $\yc$ between the corresponding parts of the action.
Without this trick, perturbation theory breaks down altogether: $\log\rate_\text{NLO} > 0$ for all the values of $x$ studied.

Though we have not done so here, the next-to-next-to-leading order (NNLO) perturbative results could be constructed from the numerical results of Refs.~\cite{Ekstedt:2021kyx, Ekstedt:2022ceo}, following the prescription for scale-shifters in Ref.~\cite{Gould:2021ccf}.

Comparing lattice and perturbation theory for the nucleation rate shows a similar trend as for the equilibrium quantities: NLO corrections are essential for any quantitative agreement, but still sizeable discrepancies remain.
At fixed $\yc-y$ and at LO in powers of $x$, perturbation theory significantly overestimates the nucleation rate: $(-\log\rate_\text{LO})/(-\log\rate) \sim 1/2$, or very roughly $\rate_\text{LO}/\rate \sim e^{60}$.
The perturbative rate is also a shallower function of $\yc-y$ over the range studied.
Conversely, if one is interested in the degree of supercooling at a fixed value of the rate, LO perturbation theory gives a significant underestimate: $(\yc -y)_\text{LO}/(\yc -y)\sim 1/2$.

Extending to NLO in powers of $x$, perturbation theory performs significantly better.
\fixes{At the two smaller values of $x$,} the logarithm of the NLO nucleation rate at fixed $\yc-y$ agrees with the lattice to within about 10-20\% over the range studied. \fixes{However by $x=0.05$, the convergence of perturbation theory is failing and NLO is no closer than LO.}
However, in all cases the slope of the rate appears to disagree significantly, implying disagreement on the duration of the transition (see Sec.~\ref{sec:cosmo}) and on the rate at both larger and smaller supercooling.

\subsection{Visualising bubble nucleation} \label{sec:bubbles}

\begin{figure}[t]
  \centering
    \includegraphics[width=0.48\textwidth]{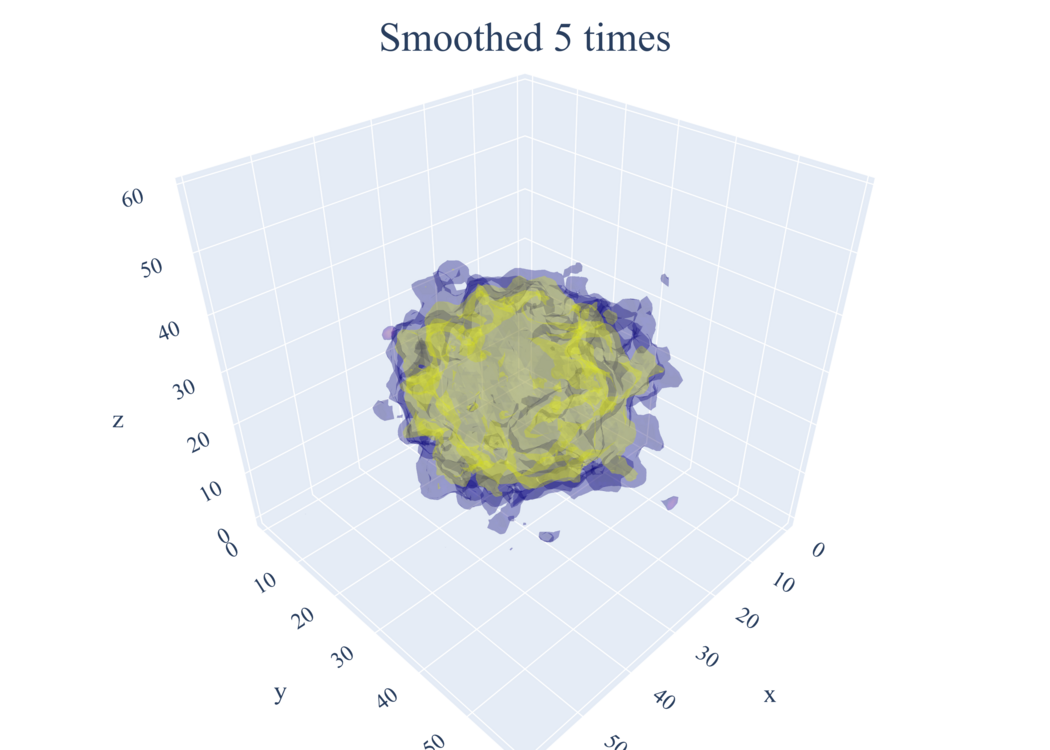}
    \includegraphics[width=0.48\textwidth]{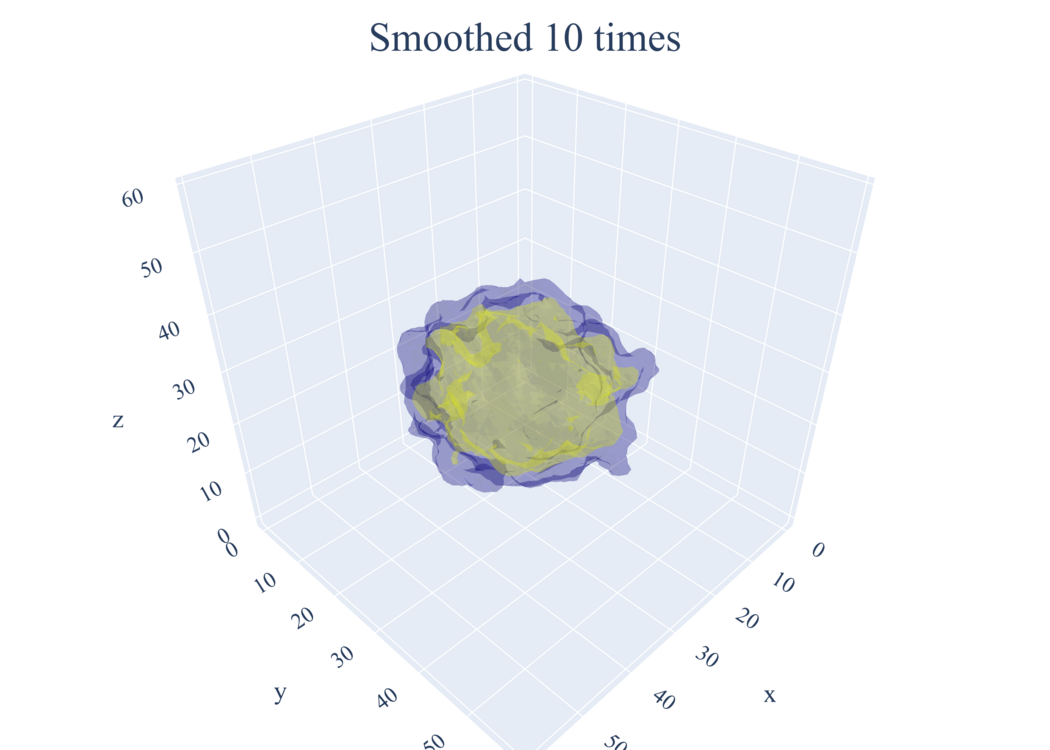}
    \caption{
    Isosurface plot of $\phi^\dagger \phi$ showing a configuration from the separatrix, from a simulation at the parameter point $x=0.02602457$, $y=0.0152473$, $g_3^2a = 4/7.332605$.
    The coordinate axes are displayed in lattice units, in which the box size is $60^3$. The bubble centre has been translated to the centre of the box.
    To remove some lattice-scale fluctuations, the Higgs field has been smoothed according to Eq.~\eqref{eq:smoothing}.}
    \label{fig:bubble}
\end{figure}

Typical lattice field configurations in the Boltzmann distribution have fluctuations on the lattice scale.
In the continuum limit $a\to 0$, these fluctuations give an infinite contribution to the free energy, and to the mass of the Higgs field, essentially a manifestation of the Ultraviolet Catastrophe.
These infinite contributions are cancelled by counterterms, which grow as $1/a$ and $\log a$ in the continuum limit.
Due to the superrenormalisability of this theory, its ultraviolet behaviour is simple, and hence the counterterms can be computed exactly \cite{Laine:1995np}.
This cancels divergences in physical observables such as the latent heat or surface tension, but not in the lattice field configurations themselves, which retain lattice-scale fluctuations.
Thus, bare lattice configurations do not have a good continuum limit, and we cannot simply identify the most likely configurations between the two phases as physical critical bubbles.

We can, however, construct a field configuration with a good continuum limit by coarse-graining over the lattice scale.
Such coarse-graining is also motivated within perturbative approaches to bubble nucleation \cite{langer1974metastable, Gould:2021ccf}: In the presence of a hierarchy of scales coarse-graining is necessary to correctly describe the critical bubble at leading order.
In radiatively-induced transitions (such as this one) coarse-graining is also necessary for the existence of the critical bubble within perturbation theory.

For our coarse-graining procedure, we smooth the Higgs field by combining the field at each point with its neighbours, suitably parallel transported,
\begin{align} \label{eq:smoothing}
\phi(x) &\to \frac{1}{4} \phi(x) + \frac{1}{8}\sum_{i}U_i(x)\phi(x+i) \nonumber\\
&\qquad + \frac{1}{8}\sum_{i}U_i^\dagger(x-i)\phi(x-i),
\end{align}
where $i$ runs over the three positive links to neighbouring lattice sites, and $U_i(x)$ refers to the lattice gauge link variables.
After smoothing the field $n$ times, structures smaller than $na$ are washed out, but larger scale structures are largely unaffected.
Thus, if one is interested in the field on physical length scales $\sim l$, this should be well exhibited by choosing $1 \ll n \ll l/a$.

Figure \ref{fig:bubble} shows two isosurface plots of $\phi^\dagger \phi$ after smoothing the Higgs field to remove lattice-scale noise.
The particular configuration was extracted from the Boltzmann distribution, saved to file as the Markov chain crossed the separatrix $\phi^\dagger\phi=\varphi^2_\text{C}$.
The origin of the coordinates has been shifted so that the bubble lies in the centre.
A video showing a sequence of isosurface plots for an example real-time trajectory of bubble growth can be seen at \cite{videos}.

For weaker transitions (larger $x$), the bubble configurations relevant for transitions with $\action\sim 100$ are comparatively larger.
The trend in bubble size can be understood from an argument based on the thin wall approximation, as a consequence of the surface tension decreasing for weaker transitions.
That is, for a bubble of fixed size, decreasing the surface tension increases the nucleation rate, and thus to return to $\action\sim 100$ one must increase the size of the bubble.

\subsection{Cosmological evolution} \label{sec:cosmo}

Given the bubble nucleation rate, one can determine the bulk evolution of the phase transition \cite{Guth:1981uk, Enqvist:1991xw, Anderson:1991zb}.
The onset of the transition, when there is approximately one bubble nucleated within a Hubble volume, occurs approximately when
\begin{align} \label{eq:one_bubble_per_hubble}
-\log \frac{\Gamma}{H^4} + \log\frac{\beta}{H} = 0,
\end{align}
where $H$ is the Hubble rate, and $\beta$ can be defined as%
\footnote{
In the semiclassical approximation, the rate can be written $\Gamma = A e^{-S}$, and the usual definition reads $\beta\equiv -dS/dt$.
Beyond the semiclassical approximation, there is a degree of arbitrariness in how one generalises this, amounting to a small uncertainty in $\beta$ of the form $d\log (A/T^4)/d\log T$.
}
\begin{align} \label{eq:beta}
\beta \equiv \frac{d}{d t} \log \frac{\Gamma}{T^4} = -HT \frac{d}{dT}\log \frac{\Gamma}{T^4},
\end{align}
equal to the inverse of the time over which $f_{\rm sym}$ varies by an $O(1)$ amount.
In the second equality of Eq.~\eqref{eq:beta}, we have made the assumption of radiation domination, so that $dT/dt = -HT$.

As the phase transition proceeds, the fraction of space which remains in the symmetric phase $f_{\rm sym}$ satisfies the following approximate equality
\begin{align} \label{eq:percolation_4d}
-\log \Gamma + \log\beta^4 - \log\frac{8\pi}{3}v_{\rm w}^3 + \log(-\log f_{\rm sym}) = 0,
\end{align}
where $v_{\rm w}$ is the bubble wall speed.
Due to the strong exponential growth of the bubble nucleation rate with time, the temperature at which the phase transition takes place is relatively well defined, independently of the precise choice of $f_{\rm sym}\sim 1$.

Rewriting Eq.~\eqref{eq:percolation_4d} in terms of the rescaled lattice nucleation rate gives the following approximate condition for percolation
\begin{align}
\label{eq:percolation}
\action + 4 \log \tilde{\beta} \approx 137,
\end{align}
where we have defined $\tilde{\beta}\equiv -\partial_y \log \rate$.
In deriving the numerical value on the right hand side, we have assumed the transition to take place at $T\approx 140$ GeV \cite{Gould:2019qek, Ramsey-Musolf:2019lsf}, and have substituted SM-like parameters in order to arrive at the numerical values \cite{Moore:2000jw, Caprini:2019egz}.
This value, 137, will vary by $O(1)$ depending on the specific 4d model considered.
We have also approximated the factor $\beta$ as
\begin{align} \label{eq:beta_4d_3d}
\beta \approx H \eta_y \cdot \tilde{\beta},
\end{align}
neglecting the parametrically slower $T$-dependence of $x$ and $g_3^2/T$, in comparison with that of $y$; see the discussion around Eq.~\eqref{eq:etas}.
Note the factorisation of ultraviolet and infrared contributions in Eq.~\eqref{eq:beta_4d_3d}.

\begin{figure}[t]
    \centering
    \includegraphics[width=0.48\textwidth]{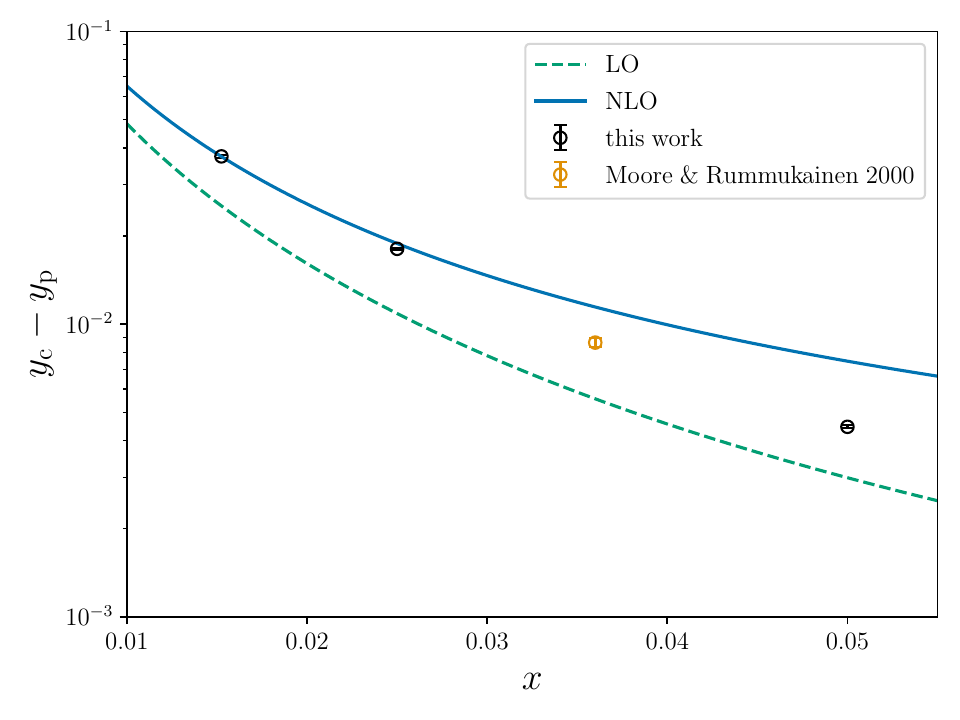}
    \caption{
    The degree of supercooling at percolation, as determined by Eq.~\eqref{eq:percolation}. 		Lattice results from this paper and from Ref.~\cite{Moore:2000jw} are shown.
	}
    \label{fig:y_cosmo}
\end{figure}

\begin{figure}[t]
    \centering
    \includegraphics[width=0.48\textwidth]{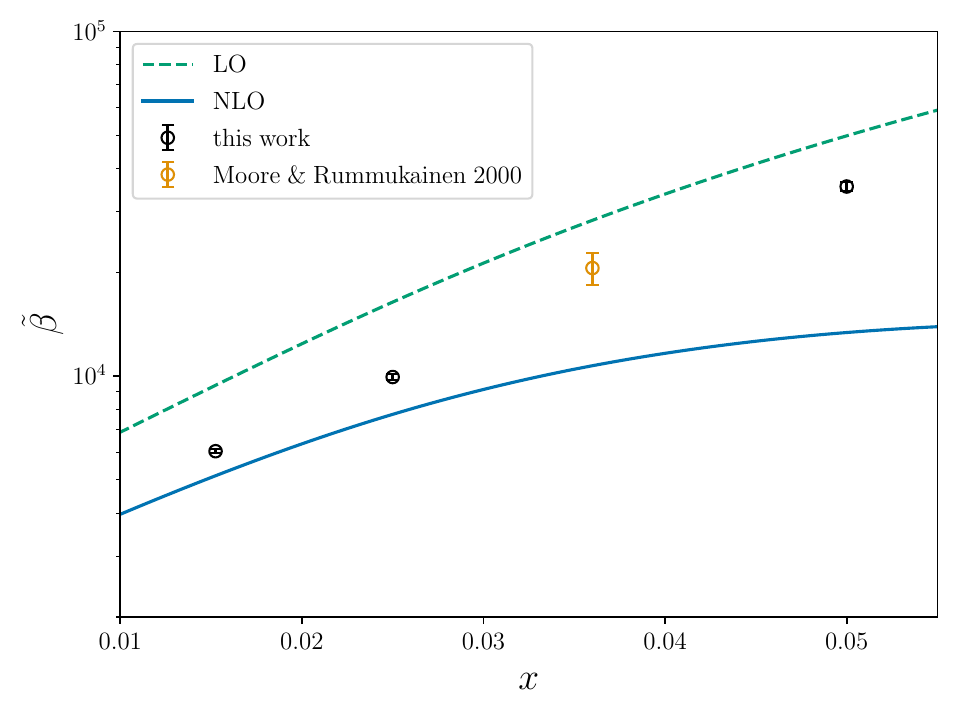}
    \caption{
    The (scaled) inverse duration of the transition, evaluated at percolation.
    Lattice results from this paper and from Ref.~\cite{Moore:2000jw} are shown.
	}
    \label{fig:beta}
\end{figure}

\begin{table}[t]
	\centering
    \begin{tabular}{lll}
      \hline
      \muco{$x$} &  \muco{$\yc-\yp$} & \muco{$\tilde{\beta}_\text{p}$} \\
      \hline
      0.0152473 & 0.03741(47) & $6.052(80)\times 10^3$ \\
      0.025 & 0.01809(16) & $9.93(18)\times 10^3$ \\
      0.036$^*$ & 0.00865(30) & $2.06(21) \times 10^4$ \\
      0.05$^*$ & 0.004461(51) & $3.547(98)\times 10^4$ \\
      \hline
    \end{tabular}
  \caption{Lattice Monte-Carlo results for the degree of supercooling and (scaled) inverse duration of the transition, from this work, and from Ref.~\cite{Moore:2000jw}.
  The asterisks $(^*)$ are a reminder that, for these values of $x$, only two lattice spacings have been used to extrapolate to $a\to 0$.
  \label{table:results_cosmo}}
\end{table}

Fig.~\ref{fig:y_cosmo} shows the degree of supercooling at percolation, $\yc - \yp$, and Fig.~\ref{fig:beta} shows the (scaled) inverse duration of the transition, $\tilde{\beta}_\text{p}$.
Together with the lattice data, we have included the LO and NLO perturbative results.
Again we have used a strict expansion in $x$ to determine the NLO values, thereby ensuring order-by-order gauge invariance.
The NLO perturbative results agree well with the lattice for the smallest value of $x$, but by $x=0.05$ are no closer than the LO results.

\section{Conclusions} \label{sec:conclusions}

In this paper, we have carried out extensive lattice Monte-Carlo simulations and continuum extrapolations of the $\mr{SU}(2)$ Higgs 3d EFT. This model describes the high temperature thermodynamics of the Standard Model electroweak sector (up to small corrections), as well as the thermodynamics of a wide range of BSM extensions of the electroweak sector. We have focused on the region of parameter space containing first-order phase transitions in this EFT, motivated by planned gravitational wave experiments, and by the possibility of successful electroweak baryogenesis.

Our results significantly extend those of the original works \cite{Kajantie:1995kf, Moore:2000jw}, and provide a data resource which can be used to reliably and quantitatively determine properties of first-order phase transitions in extensions of the electroweak sector. Due to the factorisation of infrared and ultraviolet contributions, evidenced in Eqs.~\eqref{eq:latent_heat_condensates}, \eqref{eq:sigma_T}, \eqref{eq:rate_relation} and \eqref{eq:beta_4d_3d}, once the effective couplings of this 3d EFT have been calculated for a given 4d model, the nonperturbative thermodynamics of the 4d model can be simply read off. The results for the bubble nucleation rate also allow one to estimate the gravitational wave spectrum produced by sound waves in the fluid plasma \cite{Hindmarsh:2013xza, Hindmarsh:2015qta, Hindmarsh:2017gnf}. This approach was adopted in Ref.~\cite{Gould:2019qek}, using the single lattice result for the bubble nucleation rate from Ref.~\cite{Moore:2000jw}.

The results of this paper are directly applicable to BSM scenarios in which new degrees of freedom, with effective masses at the transition of order $O(gT)$ or greater, induce the Higgs symmetry-breaking transition to be first order. However, in scenarios where BSM degrees of freedom are lighter than this, or participate directly in the transition, such as in two-step phase transitions, new lattice Monte-Carlo simulations in other 3d EFTs are necessary; see for example Refs.~\cite{Laine:1998qk, Laine:1998wi, Laine:2012jy, Kainulainen:2019kyp, Niemi:2020hto, Gould:2021dzl}, and Refs.~\cite{Cossu:2020yeg, Galati:2021njb} in a different context.

By densely scanning the parameter space of the 3d $\mr{SU}(2)$ Higgs model, we have revealed information on the  functional forms of key thermodynamic quantities, such as the critical mass and the jump in the Higgs quadratic condensate. Beyond their intrinsic value, these can be used to provide rigorous tests of different perturbative approaches \cite{Buchmuller:1994vy, York:2014ada, Ekstedt:2022zro, Ekstedt:2022ceo}, and alternative nonperturbative approaches \cite{Elias-Miro:2020qwz, Sberveglieri:2020eko}, as a function of the perturbative expansion parameter; see for example Refs.~\cite{Arnold:1996zj, Tetradis:1996kh, Gould:2021dzl}. While perturbative approaches to the thermodynamics of non-Abelian gauge theories are fundamentally incomplete \cite{Linde:1980ts}, the computational cost of lattice Monte-Carlo simulations means that complementary approaches are necessary for exploring the high dimensional parameter spaces of possible physical models. Reliably benchmarking our confidence in such complementary methods is therefore imperative.

We have tested the first two orders of the perturbative expansion in $x$, and find the NLO approximation to provide reasonable estimates for most of the quantities studied, at least for $x\lesssim 0.05$. However, further work at higher orders in this expansion is necessary to determine to what extent it converges onto the lattice results, especially given the large disparity between LO and NLO. The extension to NNLO is presented in contemporary works \cite{Ekstedt:2022zro, Ekstedt:2022ceo}.

A possible extension of this work would be to perform a similar lattice Monte-Carlo study of the 3d $\mr{U}(1)$ Higgs model. This model does not suffer from the Infrared Problem \cite{Linde:1980ts}, its infrared sector being free. Hence, by extending Refs.~\cite{Kajantie:1997hn, Kajantie:1997vc} and simulating a number of different phase transition strengths in the $\mr{U}(1)$ Higgs model, one would be able to address the question: how numerically important is the Infrared Problem?

The simulations presented in this paper made use of well-understood and highly efficient algorithms for the study of the thermodynamics of 3d bosonic theories \cite{Berg:1992qua, Kajantie:1995kf}. However, our study has revealed a need to develop more efficient algorithms specifically for the study of the bubble nucleation rate. The calculation of the bubble nucleation rate requires large lattices to comfortably fit a critical bubble, and in this case the multicanonical algorithm of Ref.~\cite{Kajantie:1995kf} requires very long Markov chains in order to sample the full phase space. The underlying reason is that the volume-averaged order parameter $\phi^\dagger\phi$ cannot distinguish between nascent, localised bubbles and delocalised fluctuations spread over the lattice, preventing the multicanonical algorithm from efficiently tunnelling between the symmetric phase and field configurations containing critical bubbles. Development of more efficient algorithms for simulating bubble nucleation is therefore an important next step.

\section*{Acknowledgements}
The authors wish to thank
A.~Ekstedt,
J.~Hirvonen,
A.~Kormu,
J.~L\"{o}fgren,
T.V.I.~Tenkanen
and D.~Weir
for enlightening discussions at various stages of this project,
and A.~Ekstedt for his insightful comments on a draft version.
\fixes{We would also like to thank
L.~Friedrich,
M.~Diaz,
M.~Ramsey-Musolf
and T.V.I.~Tenkanen
for bringing to our attention some typos in our perturbative results in the first version.}
O.G.~(ORCID ID 0000-0002-7815-3379) was supported by the Research Funds of the
University of Helsinki, and U.K.\ Science and Technology Facilities
Council (STFC) Consolidated grant ST/T000732/1.
S.G.~(ORCID ID 0000-0002-8960-1392) was supported by the Academy of Finland grant 320123.
K.R.~(ORCID ID 0000-0003-2266-4716) was supported by the Academy of Finland grants 319066, 320123 and 345070.
The authors wish to acknowledge CSC -- IT Center for Science, Finland, for generous computational resources.

\appendix

\begin{figure}[t]
    \includegraphics[width=0.48\textwidth]{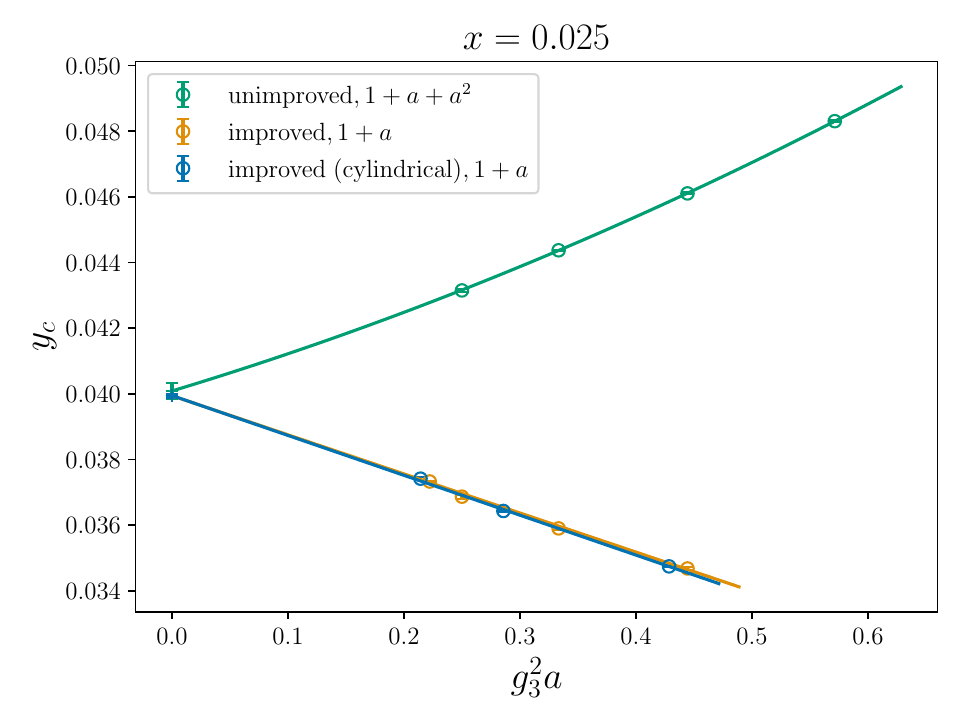}
    \includegraphics[width=0.48\textwidth]{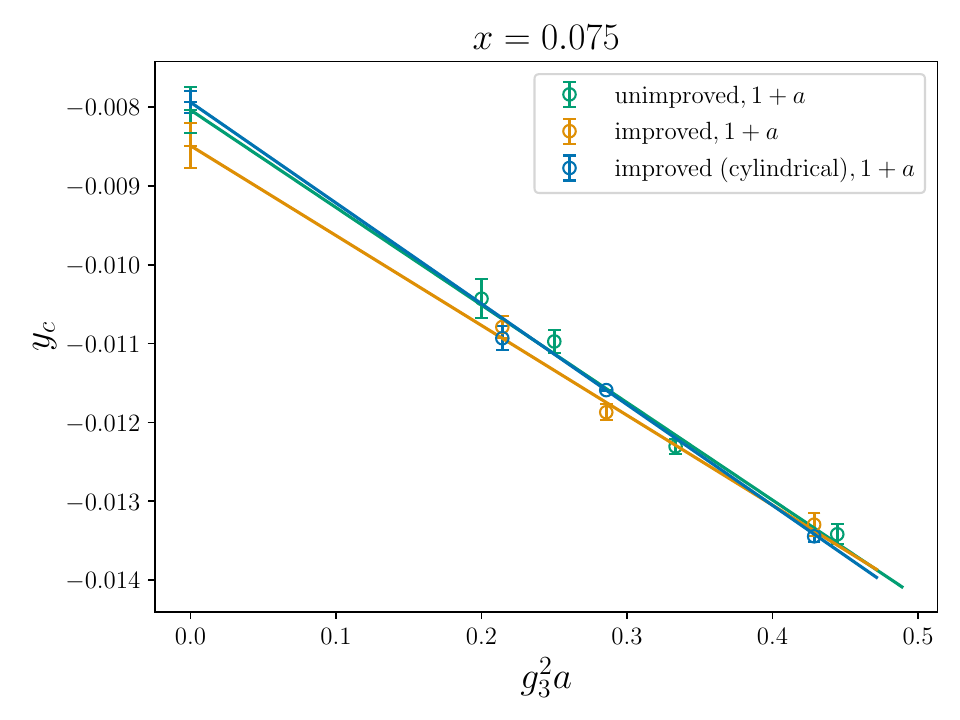}
    \caption{
      Extrapolations to the continuum limit for the critical mass $\yc$ at two different values of $x$.
    }
    \label{fig:continuum_limits_yc}
\end{figure}

\begin{figure}[t]
    \includegraphics[width=0.48\textwidth]{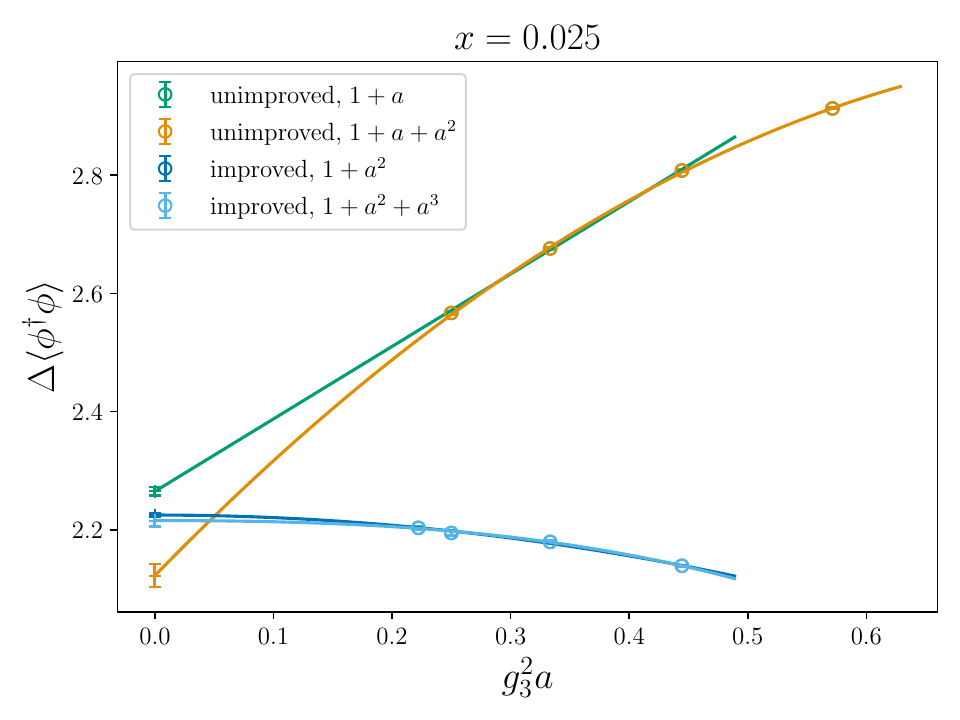}
    \includegraphics[width=0.48\textwidth]{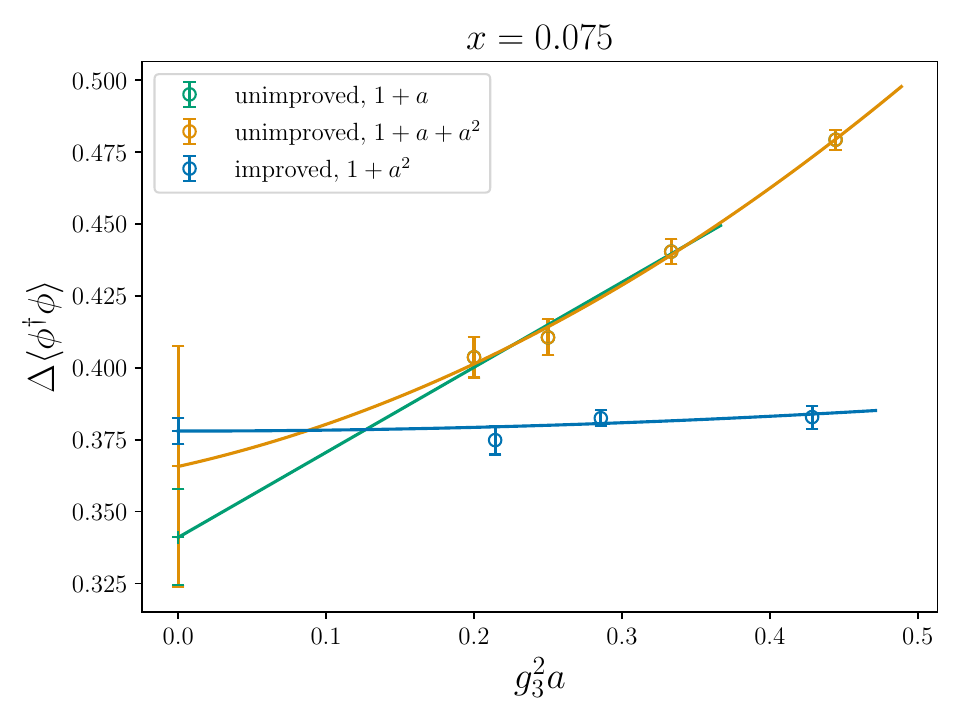}
    \caption{
      Extrapolations to the continuum limit for the Higgs quadratic condensate at two different values of $x$.
    }
    \label{fig:continuum_limits_phi2}
\end{figure}

\section{Continuum extrapolations} \label{appendix:extrapolations}
\begin{figure}[t]
    \includegraphics[width=0.48\textwidth]{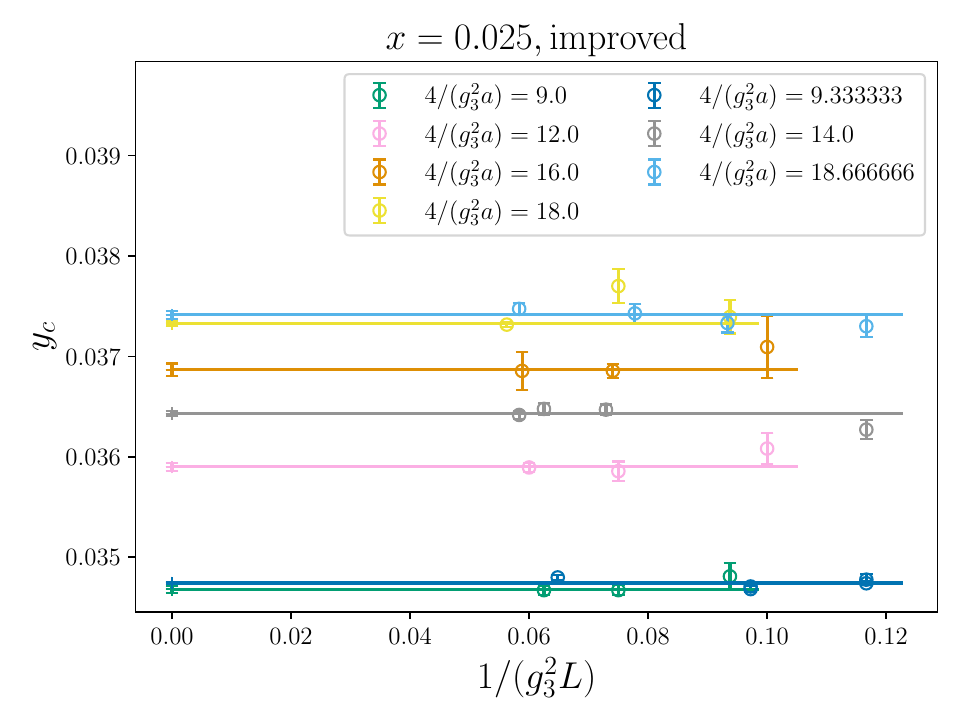}
    \includegraphics[width=0.48\textwidth]{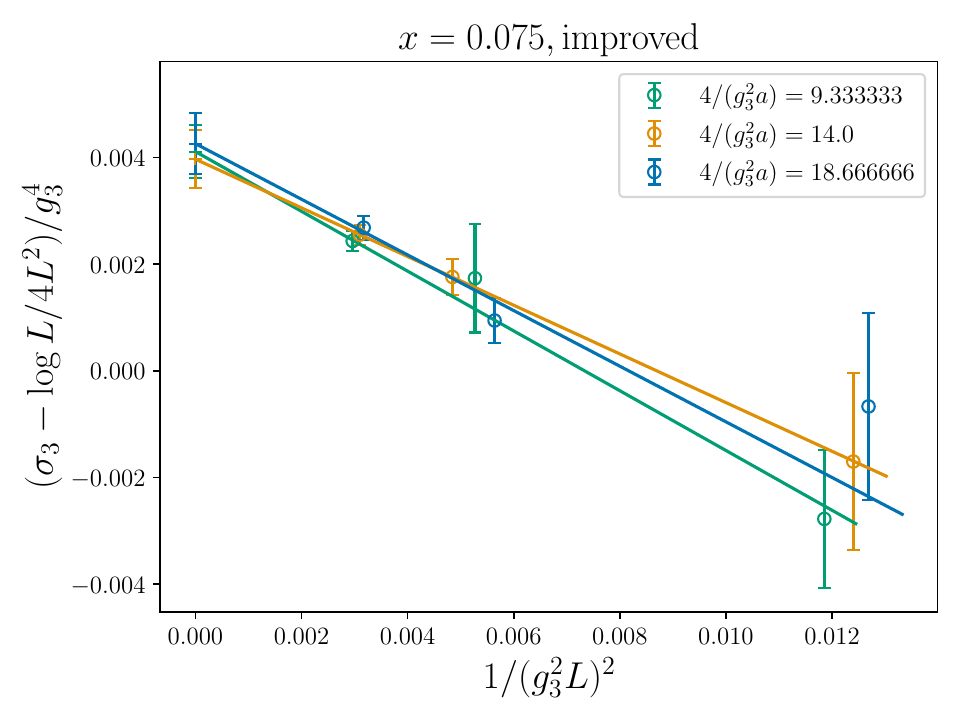}
    \caption{
      \newtext{Extrapolations to the infinite-volume limit for the critical temperature at $x=0.025$ (above), and the surface tension at $x=0.075$ (below). For the critical temperature, volume corrections are exponentially suppressed, and hence we simply fit a constant to the largest few volumes. For the surface tension, the extrapolation of the surface tension subtracts off the known contributions of capillary waves, Eq.~\eqref{eq:surface_tension_volume}.}
    }
    \label{fig:volume_limits}
\end{figure}

\begin{figure}[t]
    \includegraphics[width=0.48\textwidth]{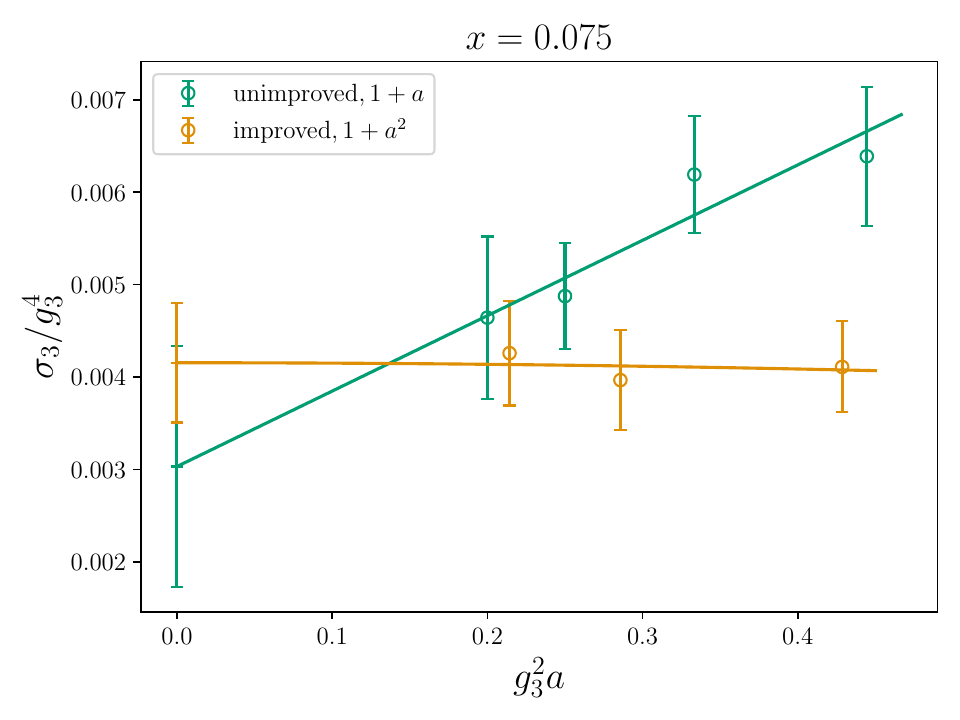}
    \includegraphics[width=0.48\textwidth]{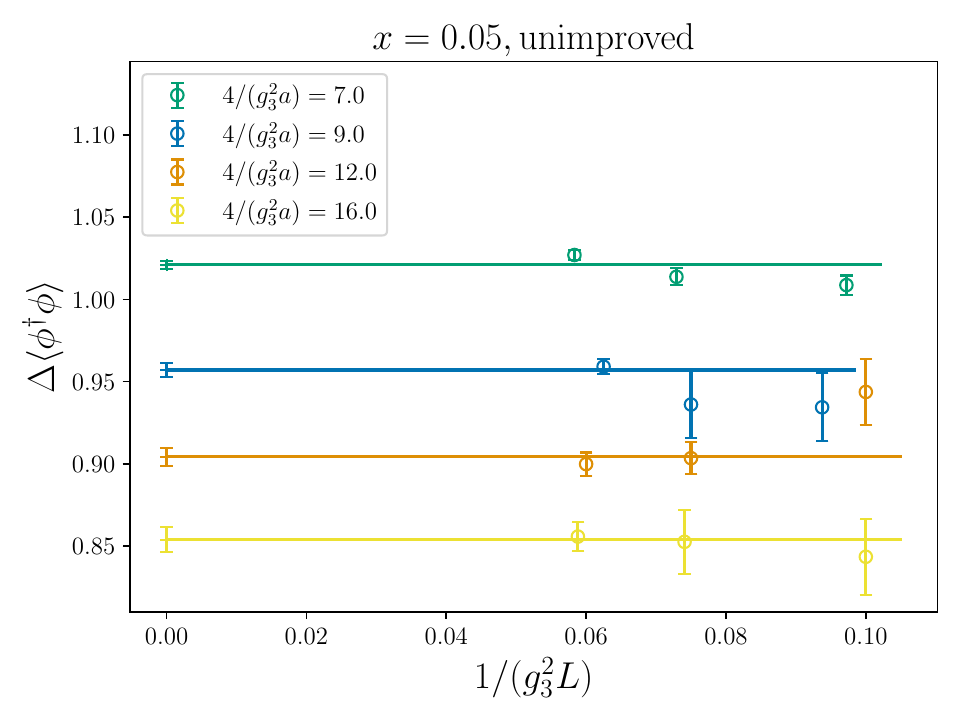}
    \caption{
      \newtext{The figure above shows the extrapolation to the continuum limit for the surface tension at $x=0.075$. The figure below shows the extrapolation to the infinite volume limit for the discontinuity in the quadratic Higgs condensate. For the latter, volume corrections are exponentially suppressed, and hence we simply fit a constant to the largest few volumes.}
    }
    \label{fig:continuum_limit_sigma}
\end{figure}

\begin{figure*}[t]
    \centering
    \includegraphics[width=0.48\textwidth]{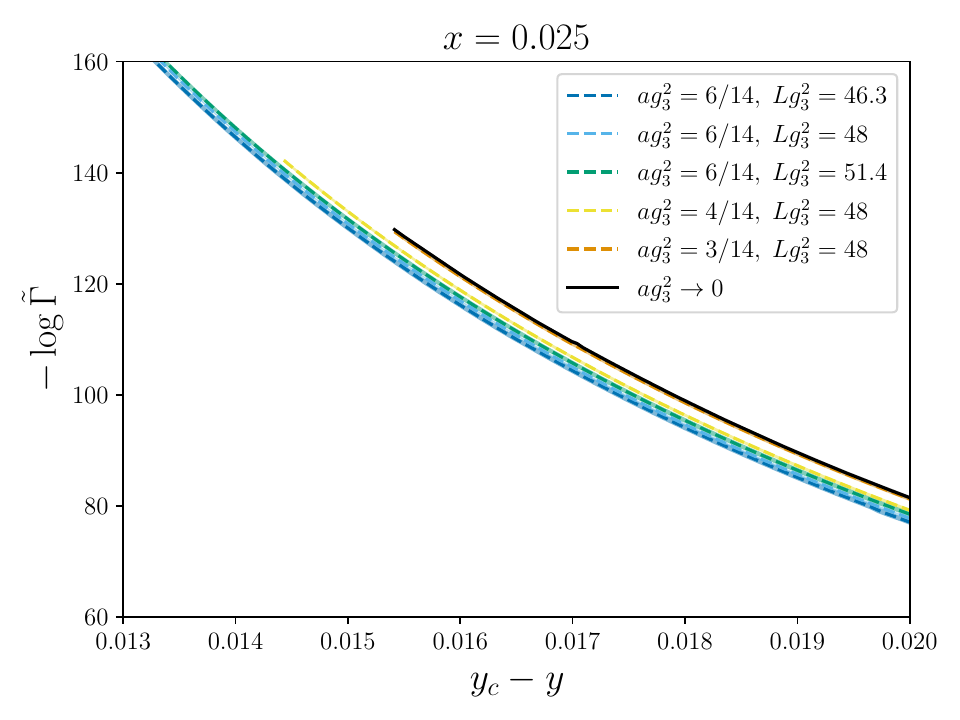}
    \includegraphics[width=0.48\textwidth]{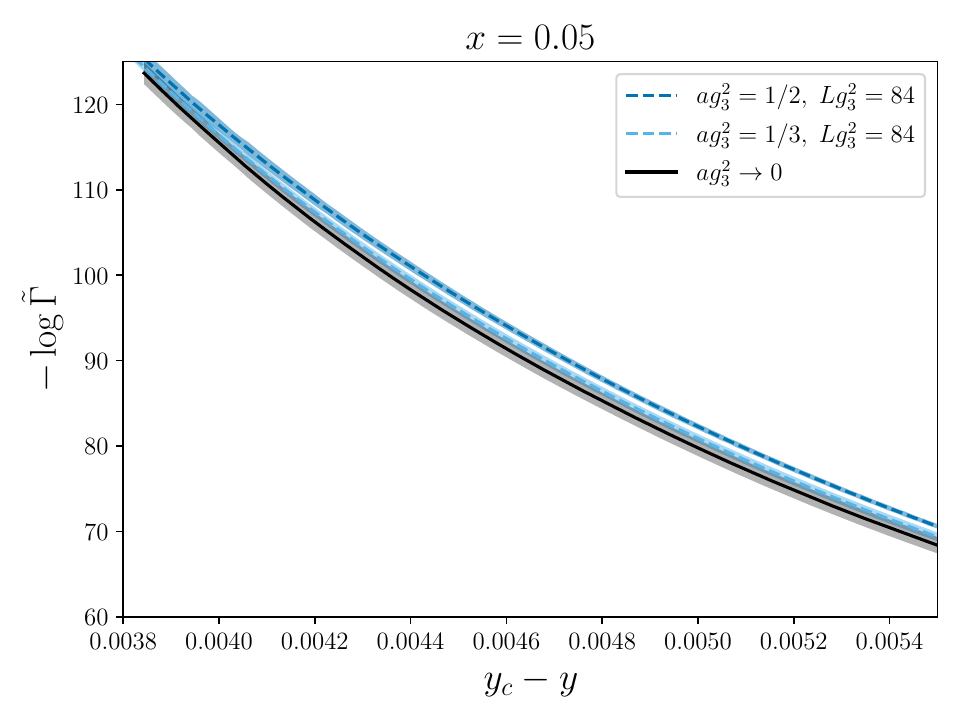}
    \caption{
    Extrapolations to the continuum limit for the nucleation rate at two different values of $x$.
    }
    \label{fig:continuum_limits_rate}
\end{figure*}

\newtext{In this appendix, we collect together a representative set of plots showing our continuum extrapolations of lattice data on finite lattices; Figs.~\ref{fig:continuum_limits_yc}, \ref{fig:continuum_limits_phi2}, \ref{fig:volume_limits}, \ref{fig:continuum_limit_sigma} and \ref{fig:continuum_limits_rate}. As discussed in Secs.~\ref{sec:critical_temperature}, \ref{sec:latent_heat}, \ref{sec:surface_tension} and \ref{sec:nucleation_rate}, the volume and lattice-spacing dependence differs depending on the observable.}

\newtext{Observables depending only on bulk quantities in homogeneous phases, $\yc$, $\Delta \langle \phi^\dagger \phi\rangle_{\rm c}$ and $\Delta \langle (\phi^\dagger \phi)^2 \rangle_{\rm c}$, have only exponentially small volume dependence in sufficiently large volumes. The same is true for the bubble nucleation rate on lattices where the bubble takes up a sufficiently small fraction of the lattice, such that it cannot see itself through the periodic boundaries. In these cases, we simply fit a constant to the results from the largest few lattices, ensuring that they agree within error. The lattice spacing of the surface tension is more complicated, but the leading dependence is known analytically so can be subtracted off, and the remaining dependence fit to.}

\newtext{Without the $O(a)$ improvement, all the physical quantities that we consider suffer from $O(a)$ corrections at finite lattice spacing. In all these cases we have performed polynomial fits, choosing the lowest degree polynomial such that $\chi^2/\text{d.o.f.}\sim 1$, either $1+a$ or $1+a+a^2$. With the $O(a)$ improvement, the linear lattice spacing dependence of all quantities except $\yc$ is cancelled, so the linear term is omitted from the polynomial fits.}
The benefits of the $O(a)$ improvement for the condensates \newtext{and surface tension} are marked.

\section{Perturbative results} \label{appendix:perturbative_results}

Here we collect some perturbative results for the 3d SU(2) Higgs theory. The beta functions, $\beta_{\kappa} \equiv d\kappa/d\log\mu_3$, are \cite{Farakos:1994kx}
\begin{align}
\beta_y &= \frac{1}{(4\pi)^2}\left(-\frac{51}{16} - 9 x + 12 x^2\right), \label{eq:running_y} \\
\beta_x &= \beta_{g_3^2} = 0.
\end{align}
These are exact, due to the superrenormalisability of the theory.

At tree-level in the 3d EFT, symmetry-breaking phase transitions occur at $m_3^2=0$ and appear to be of second order.
However, loop corrections may modify the order of the transition, and for a first-order phase transition the critical temperature occurs for $m_3^2>0$.
A strict loop expansion for the transition leads to infrared divergences and stray imaginary parts at $O(\hbar^2)$ \cite{Laine:1994zq}, demonstrating the inapplicability of the loop expansion in this case.

For sufficiently small $x$, a perturbative expansion in $x$ can be constructed \cite{Kajantie:1997tt, Kajantie:1997hn, Moore:2000jw}. Doing so requires including some one-loop terms into the LO approximation, thereby resumming the loop expansion. The resulting expansion is closely related to the coupling expansion for the case where the 4d couplings satisfy $\lambda\sim g^3$ \cite{Arnold:1992rz, Ekstedt:2020abj}. In contemporary works, this expansion has been formalised and extended to NNLO \cite{Ekstedt:2022zro, Ekstedt:2022ceo}. 

The EFT approach to thermal bubble nucleation \cite{Gould:2021ccf} offers a consistent method to calculate the nucleation rate in perturbation theory.
This approach is based on the construction of the nucleation scale effective action $S_\text{nucl}$, by integrating out all parametrically heavier modes; which can be carried out directly at the level of the path integral \cite{Hirvonen:2022jba}.
In the context of the SU(2) Higgs theory at small $x$, the gauge fields can become parametrically heavier than the Higgs field on the bubble, and therefore must be integrated out.
The gauge fields are however light in the symmetric phase; they are scale-shifters \cite{Gould:2021ccf}.
Integrating out the gauge fields yields terms in the effective action which are local at LO and NLO, but nonlocal at higher orders.

At LO, the nucleation scale effective action is
\begin{align*}
S^\text{LO}_\text{nucl} = \int d^3 x \left[
\frac{1}{2}(\partial_i \phi)^2
+\frac{1}{2}m_3^2\phi^2 - \frac{g_3^3|\phi|^3}{4(4\pi)} + \frac{1}{4}\lambda_3\phi^4
\right].
\end{align*}
which is of $O(x^{-3/2})$ when evaluated on the critical bubble, assuming the mass is not parametrically smaller than the critical mass. Here $\phi$ is a real-valued background field.

The NLO corrections to the nucleation scale effective action are
\begin{align*}
S^\text{NLO}_\text{nucl} &= \int d^3 x \bigg[
\frac{1}{2}\left(-\frac{11g_3}{4(4\pi) |\phi|}\right)(\partial_i \phi)^2\\
&\quad +\frac{g_3^4 \phi ^2}{(4 \pi) ^2}\left(\frac{51}{32} \log \left(\frac{\mu_3 }{g_3 |\phi| }\right)+\frac{33}{64}-\frac{63}{32} \log \left(\frac{3}{2}\right)\right)
\bigg].
\end{align*}
which is of $O(x^{-1/2})$ when evaluated on the critical bubble.
This nucleation scale effective action has been discussed in Refs.~\cite{Moore:2000jw, Ekstedt:2021kyx, Gould:2021ccf, Hirvonen:2021zej, Lofgren:2021ogg}.

Corrections to $S^\text{LO}_\text{nucl}+S^\text{NLO}_\text{nucl}$ arise at $O(x^0)$ due to nucleation scale, or lighter, modes fluctuating in the bubble background.
This holds as long as the mass is not parametrically smaller than the critical mass.
These corrections depend on the shape of the critical bubble, and make up the statistical prefactor of the nucleation rate, which for this theory has been calculated in Ref.~\cite{Ekstedt:2021kyx, Ekstedt:2022ceo}.
At this same order, the nucleation rate receives corrections related to the real-time growth of the critical bubble, the dynamical prefactor \cite{Langer:1969bc, Gould:2021ccf, Ekstedt:2022tqk}.
We have stopped short of computing these corrections in our perturbative analysis.

The nucleation scale effective action can also be used to calculate the equilibrium properties of the phase transition.
This is no accident, and follows because the largest contributions to the change in free energy $\Delta F$ are those which are due to the heaviest modes, and hence $\Delta F \approx T S_\text{nucl}$ at LO and NLO.
The perturbative results for equilibrium quantities are
\begin{align}
    \yc &= \frac{
        1 - \frac{51}{2} x \log \tilde{\mu_3}
    }{2 (8 \pi)^2 x}, \\
    \Delta\langle\phi^\dagger\phi\rangle_{\rm c} &= \frac{
        1 + \frac{51}{2} x
    }{2 (8 \pi x) ^2}, \\
    \Delta\langle(\phi^\dagger\phi)^2\rangle_{\rm c} &= \frac{
        1 + 51 x
    }{4 (8\pi x)^4},\\
    \sigma_3 &=\frac{1 +\left(\frac{51 \pi ^2}{4}-\frac{339}{4}\right)x}{6 \sqrt{2} (8\pi) ^3 x^{5/2}}.
\end{align}
where \fixes{$\tilde{\mu}_3\equiv e^{\tfrac{11}{34}-\tfrac{42}{34}\log\tfrac{3}{2}} (8\pi x \mu_3) \approx 0.84 (8 \pi x \mu_3)$}.
The expressions are accurate up to $O(x^{3/2})$ in the numerators.
Here we have performed a strict expansion in $x$, which ensures order-by-order gauge invariance \cite{Hirvonen:2021zej, Lofgren:2021ogg}.
The $\mu_3$ dependence of $\yc$ ensures that $\yc-y$ is renormalisation group invariant at this order; see Eq.~\eqref{eq:running_y}.

\section{Lattice volumes} \label{appendix:lattice_volumes}
Lattice volumes used for the computation of equilibrium thermodynamics are listed in Table \ref{table:equilibrium_lattices}, and results for each lattice are collected in \cite{data}.


\begin{table}[h]
	\centering
    \begin{tabular}{llc}
      \hline
      \muco{$x$} & \muco{$a g_3^2$} & \muco{volumes$/a^3$} \\
      \hline
      0.0152473 & 0.5455087$^*$ & $16^3$, $24^3$, $32^3$, $16^2\times 80$\\
      ~ & ~ & $24^2\times 120$, $32^2\times 160$ \\
      ~ & 0.3636364$^*$  & $24^3$, $32^3$, $48^3$, $24^2\times 120$ \\
      ~ & ~ & $32^2\times 160$ , $48^2\times 240$ \\
      ~ & 0.2727273$^*$  & $32^3$, $48^3$, $64^3$, $32^2\times 160$ \\
      ~ & ~ & $48^2\times 240$ , $64^2\times 320$ \\
      ~ & 0.2181818$^*$  & $40^3$, $60^3$, $80^3$, $40^2\times 300$ \\
      ~ & ~ & $60^2\times 400$ , $80^2\times 600$ \\
      0.025 & 0.5714286  & $18^3$, $24^3$, $30^3$ \\
      ~ & 0.4444444  & $24^3$, $30^3$, $36^3$ \\
      ~ & 0.4444444$^*$  & $24^3$, $30^3$, $36^3$ \\
      ~ & 0.4285714$^*$ & $20^2\times 100$, $20^2\times 140$ \\
      ~ & ~  & $24^2\times 120$, $24^2\times 144$ \\
      ~ & ~  & $36^2\times 360$ \\
      ~ & 0.3333333  & $30^3$, $40^3$, $50^3$ \\
      ~ & 0.3333333$^*$  & $30^3$, $40^3$, $50^3$ \\
      ~ & 0.2857143$^*$  & $30^2\times 180$, $48^2\times 288$ \\
      ~ & ~  & $56^2\times 336$, $60^2\times 360$ \\
      ~ & 0.25  & $40^3$, $54^3$, $68^3$ \\
      ~ & 0.25$^*$  & $40^3$, $54^3$, $68^3$ \\
      ~ & 0.2222222$^*$ & $48^3$, $60^3$, $80^3$ \\
      ~ & 0.2142857$^*$  & $40^2\times 240$, $50^2\times 300$ \\
      ~ & ~  & $60^2\times 360$, $80^2\times 480$ \\
      0.036 & 0.5714286  & $18^3$, $24^3$, $30^3$ \\
      ~ & 0.4444444  & $24^3$, $30^3$, $36^3$ \\
      ~ & 0.3333333  & $30^3$, $40^3$, $50^3$ \\
      ~ & 0.25  & $40^3$, $54^3$, $68^3$ \\
      0.05 & 0.5714286  & $18^3$, $24^3$, $30^3$ \\
      ~ & 0.5$^*$  & $18^3$, $24^3$, $30^3$, $36^3$, $42^3$ \\
      ~ & 0.4444444  & $24^3$, $30^3$, $36^3$ \\
      ~ & 0.3333333  & $30^3$, $40^3$, $50^3$ \\
      ~ & 0.3333333$^*$  & $30^3$, $36^3$, $48^3$, $54^3$, $60^3$ \\
      ~ & 0.25  & $40^3$, $54^3$, $68^3$ \\
      ~ & 0.25$^*$  & $32^3$, $40^3$, $48^3$, $64^3$, $72^3$ \\
      0.075 & 0.4444444  & $24^3$, $30^3$, $36^3$ \\
      ~ & 0.4285714$^*$ & $20^3$, $30^3$, $40^3$, $20^2\times 120$ \\
      ~ & ~ &  $30^2\times 180$, $40^2\times 240$ \\
      ~ & 0.3333333 & $30^3$, $40^3$, $50^3$ \\
      ~ & 0.2857143$^*$ & $30^3$, $48^3$, $60^3$, $30^2\times 180$ \\
      ~ & ~ &  $40^2\times 240$, $50^2\times 300$ \\
      ~ & 0.25 & $40^3$, $54^3$, $68^3$ \\ 
      ~ & 0.2142857$^*$ & $40^3$, $60^3$, $80^3$, $40^2\times 240$ \\
      ~ & ~ &  $50^2\times 300$, $60^2\times 360$ \\
      ~ & 0.2  & $50^3$, $68^3$, $84^3$ \\
      \hline
    \end{tabular}
  \caption{\newtext{Lattices used for the simulations of equilibrium thermodynamics. Asterisks mark lattices where $O(a)$ improvement was used.}
  \label{table:equilibrium_lattices}}
\end{table}

\bibliography{refs.bib}

\end{document}